\documentclass[twoside,12pt]{article}
\usepackage{wrapfig,wrapft,graphicx,epsfig}

\def\q{\vec q}

\def\tl{\tilde l}
\def\N{${\cal N}\,\,$}

\newcommand {\bea}{\begin{eqnarray}}
\newcommand {\eea}{\end{eqnarray}}
\newcommand {\be}{\begin{eqnarray}}
\newcommand {\ee}{\end{eqnarray}}

\newcommand{\ba}{\begin{array}}
\newcommand{\ea}{\end{array}}

\def\q{{\bf q}}

\def\nf{n_{\rm f}}

\def\alphas{\alpha_{\rm s}}

\def\intd3q{
   \mu^{2\epsilon}\!\!\int {d^{3-2\epsilon}q \over (2 \pi)^{3-2\epsilon}}
}

\def\simge{\mathrel{%
   \rlap{\raise 0.511ex \hbox{$>$}}{\lower 0.511ex \hbox{$\sim$}}}}
\def\simle{\mathrel{
   \rlap{\raise 0.511ex \hbox{$<$}}{\lower 0.511ex \hbox{$\sim$}}}}

\begin{document}

\title{  Why does the Quark-Gluon Plasma at RHIC behave\\
  as a nearly ideal fluid ?}
\author{Edward Shuryak\\
Department of Physics and  Astronomy,\\
University at Stony Brook, NY 11794, USA}
\maketitle
\begin{abstract}
The lecture is a brief review of the following topics: (i) collective
flow phenomena in heavy ion collisions. The  data from RHIC  indicate robust
collective flows, well described by hydrodynamics
with expected Equation of State. The transport properties
 turned out to be unexpected, with very small viscosity; 
 (ii) physics of highly excited
 matter produced in heavy ion collisions    at $T_c<T<4T_c$ is
 different
from weakly coupled quark-gluon plasma because of
relatively strong coupling generating
 bound states
of quasiparticles;  
  (iii) wider discussion of other ``strongly coupled systems''
including strongly coupled supersymmetric
theories studied via Maldacena duality, as well as recent progress
in trapped atoms with very large scattering length.
\end{abstract}
%\eject
%\tableofcontents
\section{Introduction}

When  
 high temperature
QCD was first addressed, in 1970's (see e.g. \cite{Shu_QGP}), 
the Quark Gluon Plasma (QGP) was assumed to be a gas
 of quasiparticles (=dressed quarks and gluons) which  interact relatively
weakly with each other. All
non-perturbative results (such as lattice predictions for Equation of
State (EoS)) were not in disagreement with this view. 
The 
theoretical work continued on refining the
perturbative
calculations during   1980's and 1990's, and although it resulted in
poorly converging series for realistic $T$, the hopes 
remained to do some clever
resummation and get all the physics right perturbatively.

 If that attitude turned to be true,
 one should also be able to  model heavy ion collisions following pQCD, with
``parton cascades'' limited by some reasonable cutoff around a
 scale $\sim 1\,\mbox{GeV}$. Many advocates of this approach
did the calculations and found  very
little parton rescattering and essentially $no$ collective effects.

 However when the Relativistic 
Heavy Ion Collider (RHIC) project in Brookhaven National Laboratory
was completed and started producing data, from summer of 2000, it became
apparent that a number of robust collective effects are observed.
This implies that
the effective interaction in QGP is in fact much stronger than expected
on the basis of  the perturbation theory. 
 The data  are in fact
 in much better agreement with the predictions
based on {\em ideal  hydrodynamics},
an effective theory
based on the opposite picture of very strong interaction and very
small mean free paths\footnote{ 
We will discuss also why 
 the Equation of State calculated perturbatively is still approximately
valid.}. We will discuss below current estimates of the
{\em viscosity} of QGP, and will see that its ratio to entropy density
$\eta/s\sim 1/10$ is surprisingly small. In fact, QGP is {\em the most
perfect fluid} known, since the same ratio is  greater than 1 for most liquids
\footnote{
E.g. water, which gave its name to hydrodynamics, does not flow that well
if one considers a drop of only $\sim 1000$ molecules.}.

   Let me reformulate the question I put into the title,
{\em Why does the Quark-Gluon Plasma at RHIC behave
  as a nearly ideal fluid ?}  in a more constructive way,
by asking instead: {\em What are the QGP properties
in the strong coupling regime},
 when the gauge
field coupling   $\alphas=g^2/4\pi$ is no longer small but $\sim 1$
or even larger than that?

   Our work in this direction has started only recently, and  in this
lecture I will mostly discuss new ideas  pursued 
in our group. I.Zahed and myself \cite{SZ_newqgp} 
 suggested an explanation to hydro regime of QGP, related it
 with presence of
loosely bound pairs of quasiparticles which generate large scattering lengths.
These ideas are further developed by Brown et al \cite{BLRS}.
However this  is
just a part of developments in other
fields of physics, which happen to be quite recent as well.

One 
 revolutionary advance took place
on the theoretical front. String theorists (Maldacena) 
pointed out that one can treat  \N=4 supersymmetric gauge field
  theory in {\em
strong coupling} regime by considering a weakly coupled supergravity
problem in a particular background field. It can be done both
  at zero and finite $T$, 
and resulted in a number of fascinating (although
rather incomprehensible) results. I will briefly describe those,
as well as  recent
explanations of those, based on deeply bound
states recently suggested by I.Zahed and myself \cite{SZ_cft}.

  If this development sounds too theoretical for some readers,
here is another purely experimental one. Exciting
  recent development  took place at the frontier
of low temperature physics, with
trapped  $Li^6$ (fermionic) atoms.
 Using
 magnetic field one can use the so called Feshbach
resonances and make a pair of atoms nearly degenerate with their
bound state (usually called a molecule but actually a Cooper pair).
This  results in so large  scattering
length $a$, than a qualitatively new type of matter -- {\em strongly coupled
fermi and bose gases} -- is observed. In particular, this very dilute
systems start
to behave hydrodynamically, displaying elliptic flow very similar to that in
non-central heavy ion collisions.

Following nice traditions of the Erice schools, let me in
 these lectures  go outside nuclear physics and 
spend some time, providing 
some introductory discussion into these fascinating fields.

\section{ Collective flows and the EoS of QGP }
\subsection{ Is hadronic matter really produced in heavy ion collisions?}
Let me start with this central question, asked so often. 
Is there something qualitatively
  new in AA collisions, never
 seen in  ``elementary''\footnote{Apart of the large-$p_t$
tail, described by the parton model plus pQCD corrections, it is very far
from being elementary and is very poorly understood. One may argue that
heavy ion collisions, described well by hydro/thermodynamics, are in fact
even much
simpler. }
 pp or $e+e-$ collisions?

  Indeed, the original motivation for
 heavy ion program is not just increase the number of
 secondary particles 
produced per event (up to several thousands at RHIC), but to reach
 a {\em qualitatively different} dynamical regime.   
characterized by a small
 microscopic scale $l$ (e.g. mean free path) as compared to
the macro scale $L$ (the system's size)
\be l \ll L \ee
 If this is achieved,
 the fireball produced in
heavy ion collisions can be treated as a macroscopic
body, and
 collective ``bangs'' should be observed, rather than 
 a {\em collision-less fireworks}
 of debris created in a collision seen in $pp$ or $e^+e^-$. 
  More precisely,
if the system is macroscopically large then its description via 
{\em thermodynamics} of 
its bulk properties (like matter composition)
 and {\em hydrodynamics} for space-time  evolution should
work.

  Statistical models do indeed work remarkably well for heavy ion collisions,
as you will see from other lectures in this school. But it also
works for  $pp$ or $e^+e^-$ (and we still do not know why). 
Many years ago, in a prehistoric era before QCD, 
I have tried to
apply hydro to  $e^+e^-$: discovery of jets several years later put the
end to it. Also in 1970's, a search for signs of transverse flow
in the early hadronic spectra  at CERN ISR from pp
collisions was performed by (my former student) Zhirov and myself
 \cite{SZ_radialflow} have found
little or no signs of the collective transverse flow.
 Apparent lack of collectivity in  $pp$ or $e^+e^-$ 
 shows that although they also produce a multi-body
excited systems, those are $not$ macroscopically large.
These collisions do not produce ``matter'', just a bunch of 
outgoing particles.

The heavy ion collisions, on the other hand, show collective new
 phenomena,
 a variety of 
  {\em``flows"} to be discussed. These various forms of
 a collective expansion confirm that we do see a macroscopic behavior,
which can indeed be described using bulk properties of ``matter''. 
 Although similar ``flows''
are observed at all energies, at AGS and below they have
somewhat different origin, and I would not discuss those for lack of time.
I will  discuss mostly
 recent data from RHIC, which  prove the point 
 in a  spectacular way.

\subsection{The transverse flow }
   Let me start with historic comments. The   first  application of hydrodynamics   for  description of hadronic
fireballs   
  was suggested in the  classical  work  by  L.D.Landau 50 years ago
 \cite{landau_53}, two years after the pioneer
work on statistical model by Fermi.
 Landau has aimed his hydrodynamical theory mainly at description
of the longitudinal momenta. However
with the discovery of asymptotic freedom,
  Fermi's idea of complete stopping --the initial condition in form
of Lorentz contracted 
 disk of equilibrated matter at rest -- became problematic
\footnote{In fact Landau has been motivated
  by the $opposite$ behavior of the effective charge
in QED and scalar theories, growing toward small distances, which was 
before the asymptotic freedom  considered
the only possibility.
}. Actually we do not yet know the pre-hydro initial stage well even
  now,
so longitudinal hydro has little predictive power.

 First attempts to connect the experimental information with
   the collective transverse flow  were
 made independently  by Siemens and
   Rasmussen
\cite{SieRas} for low energy (BEVALAC) and by 
Zhirov  and  myself \cite{SZ_radialflow} for high energy pp collisions
at CERN ISR.
The idea was exactly the same: 
the collective velocity of explosion boosts 
    spectra  of  light  and  heavy
 particles differently. 
 Pions are ultrarelativistic,
and their  thermal spectrum is exponential in $p_t$: being
  boosted by flow
it    remains   exponential    with  a
 ``blue shifted''  temperature $T^*=T\gamma_t$.
     For heavy particles the effect  is  quite
 different. In particular, an infinitely heavy particle has
 no thermal velocity and its thermal spectrum
is $\sim \delta(\vec p)$. If boosted by the flow, this delta functions just
 shifts its argument by $mv_t$, where $v_t$ is the flow velocity. 
The
real nucleons or deuterons are in between of the two limits.

The findings of these two papers were however completely
different. BEVALAC spectra discussed by 
 Siemens and Rasmussen \cite{SieRas}, for heavy ions at  $E\sim 1 GeV*A$, 
 indeed have shown the expected difference for  pion  and
  proton. They were well fitted  with two
  parameters, the freezeout temperature $T_f\sim 30 \, MeV$
and the velocity of what they have called the  "blast  wave"
$v\approx  0.3$\footnote{Long discussion of whether hydro is applicable
to derive it had followed: we would not go into that now.}.
  Conclusions of our paper \cite{SZ_radialflow}  was negative:
the  $\pi,K.N$ spectra from $pp$ collisions showed a very good $m_t$-scaling 
\be {dN\over dp_t^2}\sim exp\left(-{m_t\over T}\right), \qquad
 m_t^2=p_t^2+m^2
 \ee
with the same universal slope,
without any sign of  transverse collective
 flow. So, we all  had to wait for heavy ion
collisions at similar energies
\footnote{Incidentally, let me remind that around 1980-1982
it looked quite possible that CERN ISR would be used for that purpose.
Unfortunately CERN leadership of the time decided otherwise, and even
destroyed that steady and reliable collider ``to store LEP magnets
in the building''. The extra irony of that announcement
 was the fact that LEP magnets have  low field  and
were made mostly of concrete, with small admixture of iron, so
there was no problem of their storage.  
}. 

\begin{figure}[ht]
\begin{minipage}[c]{8.cm}
 \centering 
\includegraphics[width=8.cm]{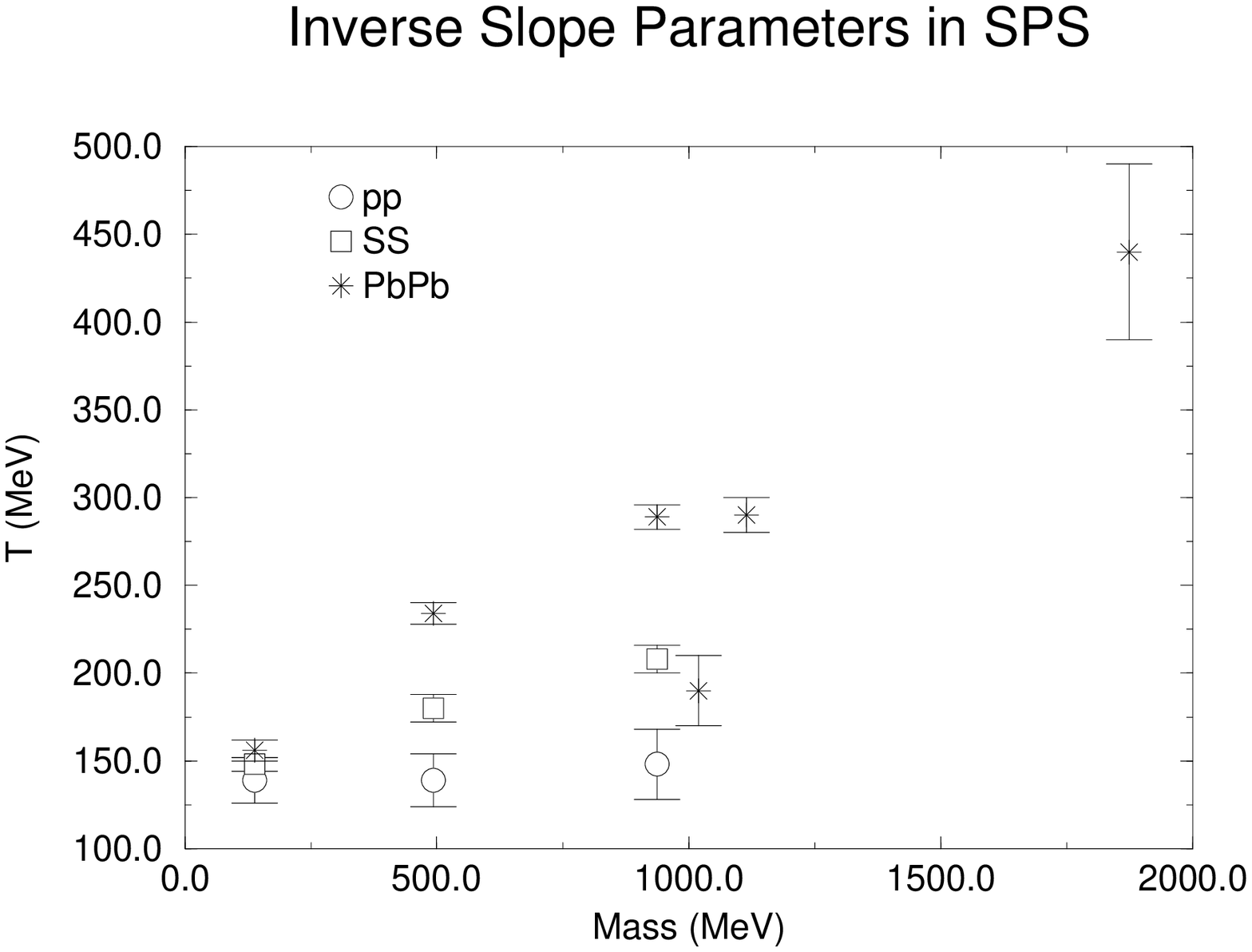} 
\end{minipage}
\begin{minipage}[c]{7.cm}
\vskip -.3cm  \includegraphics[height=6.0cm, width=6.0cm]{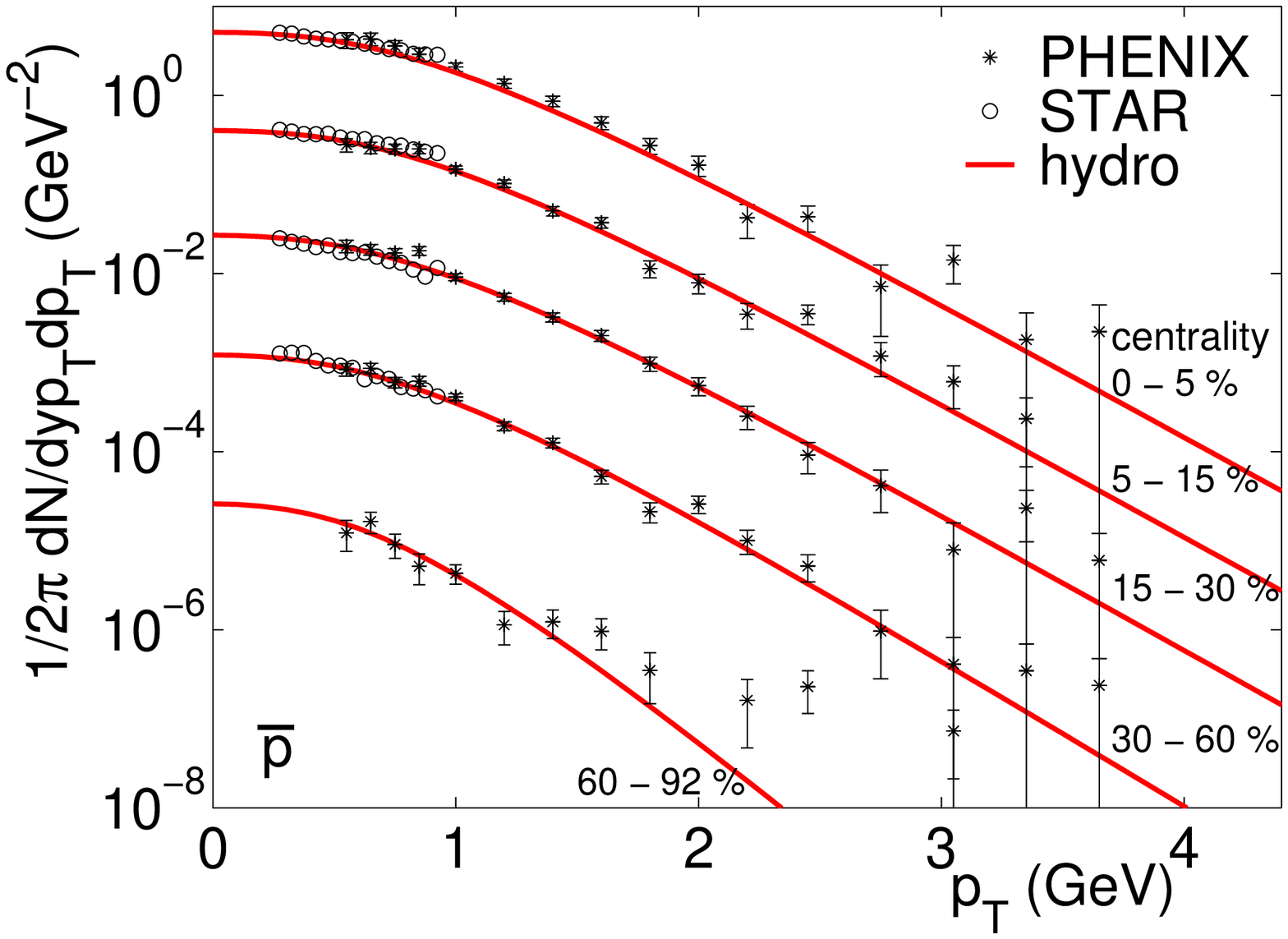}
\end{minipage}
\caption{
(a) Compilation of slopes of the $m_t$ spectra from pp collisions (ISR)
 (open circles), SS and PbPb collisions at SPS. 
(b) Comparison between STAR and PHENIX data for protons with hydro
calculation by Kolb and Rapp \cite{Kolb:2002ve} (which correctly incorporates
chemical freezeout).
\label{fig_slopes}
}
\end{figure}

It did indeed happened as planned, only  decades later, at SPS and
 at RHIC.  Fig.\ref{fig_slopes}(a), known as ``Nu Xu plot'', 
compares the $pp$ data just discussed (open circles) with slopes of
heavy ion spectra: as
one can see these slopes grows approximately linearly with 
particle mass.

  The last $d$ point is especially convincing:
if flow interpretation is
wrong and   different $m_t$ slopes in AA
and pp collisions observed are due to something else (e.g. to
 ``initial state'' rescattering, as some authors suggested), 
 the $d$ slope of the  $m_t$ spectra would be a convolution of 2 $N$ 
spectra. A simple back-on-the-envelope calculation 
proves that for
 two uncorrelated 
 nucleons the slope for $d$  should simply be
{\em the same} as for $N$.
 This however is
 contrary to experimental observations: in fact  the $d$ slope is
  about
twice larger than $p$ one.
 Only pre-existing
correlation between momenta of the two nucleons can explain that, and its
magnitude is exactly as reproduced by the flow.

   Let us now jump years ahead and show more modern version of
the same argument. Fig.\ref{fig_slopes}(b) shows recent RHIC data for the
proton $p_t$ spectra, compared with hydro prediction. Note that 
no parameters other than fitted for pion spectra are present,
and the agreement is very good, both in normalization and shape.

\begin{figure}[t]
\centering
\includegraphics[width=12.cm]{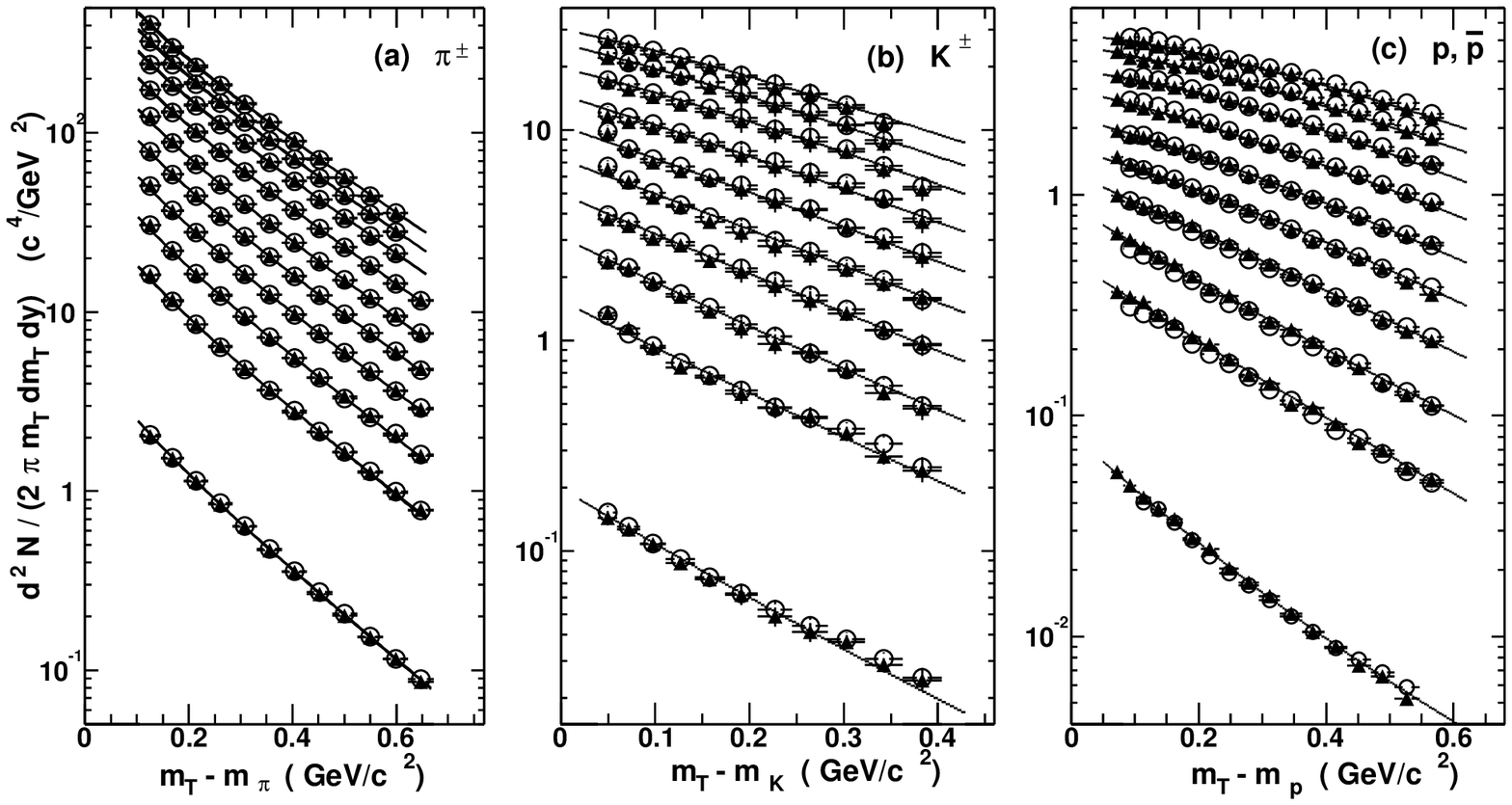}\\
\includegraphics[width=10.cm]{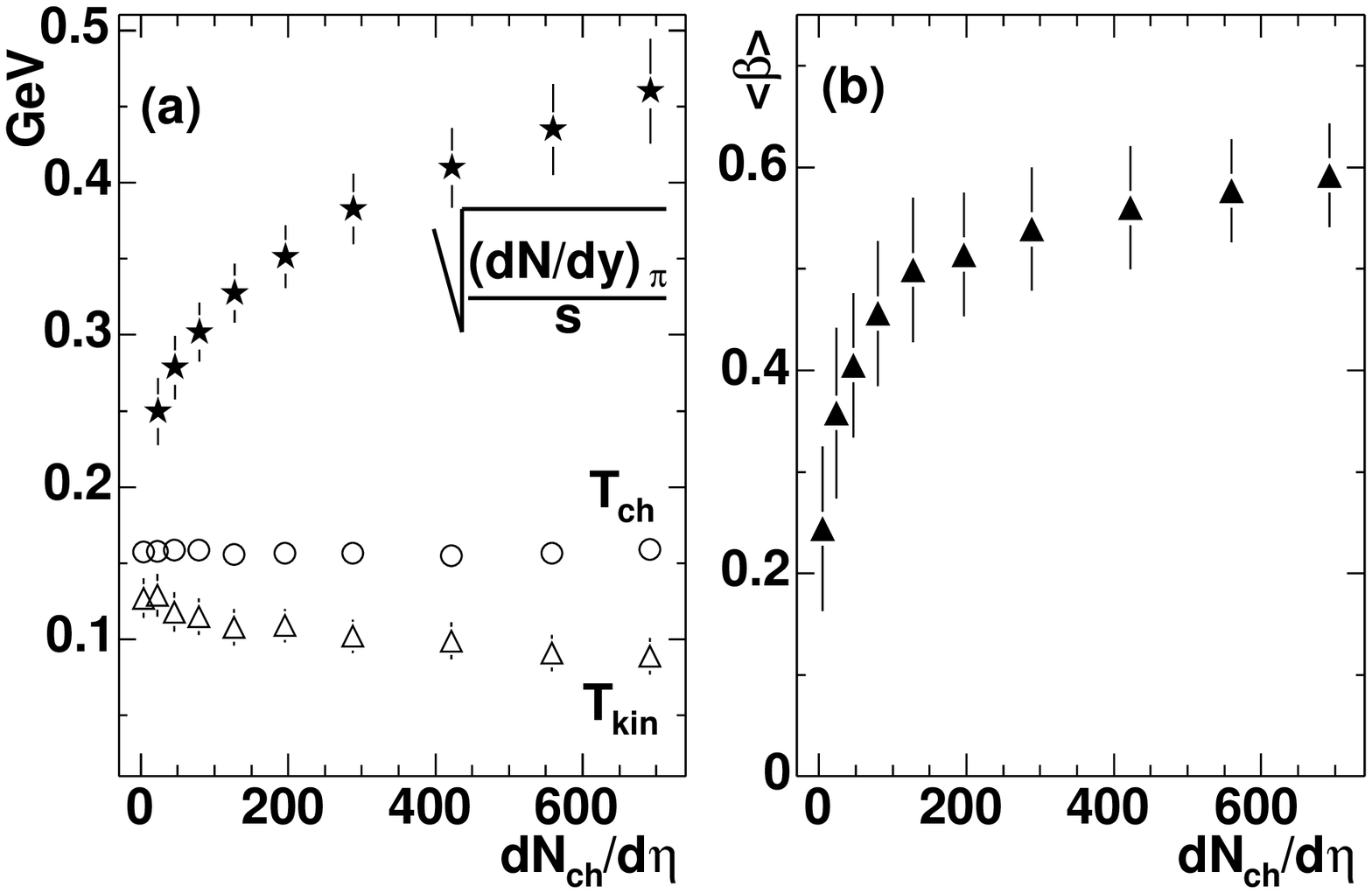}
 \caption{  \label{fig_mtspec_all}
 Pion, kaon and nucleon spectra from STAR collaboration (upper
 panels),
 together with the 
``blast model'' fits. The values of the  
and freezeout temperatures are shown in (a) and the mean collective
 velocity
in (b) part of the lower panel.}
\end{figure}

In Fig.\ref{fig_mtspec_all} we show a
compilation
of the results from RHIC (the STAR experiment) for pions, kaons and protons,
fitted with hydro-inspired ``blast wave''
parameterization. It has two basic parameters -- the freezeout
temperature $T_{kin}$ and the mean flow velocity $<\beta>$. 
Different spectra are for different centrality classes, 
central at the top, the lowest curve
 in each case corresponds to $pp$ collisions.
One can see that modification of the shapes of these
spectra of all secondaries are nicely explained by the model.
The parameter values, shown in the lower figures, show smooth variation
with the centrality: $T_{kin}$ decreases and the velocity increases. 
The reason for that was suggested by C.M.Hung
and myself \cite{HS_softest}: {\em larger systems cool
further}. So,  as
 the proverbial ``rocket scientists'', 
the most central
collisions provide the strongest conversion of the internal energy into flow. 

Note also that the temperature of chemical equilibration $T_{ch}$ 
seem to be completely independent of the 
 centrality: the interpretation of it
is that it is in fact the {\em QCD critical temperature}. For a reason which
is not yet quite clear, most inelastic reactions seem to be rapidly quenched
as soon as the matter is out of the QGP phase. 

At
 RHIC 
not only
 a strong  transverse flow with $v_t\sim .7$ was found, but 
(rather unexpectedly) hydro predictions with different slopes
hold till rather large $p_t\sim 2 \, GeV$. As a result, 
a crossing between pion and baryon spectra has been seen for the first time,
so that there are as many nucleon as pions at and above  such $p_t$,
till
$p_t\sim 5\, GeV$.

RHIC data have shown that strange quarks should flow together with
the flow, as $\phi,\Omega$ made of strange quarks show similar behavior.
One  interesting questions to be addressed soon is whether the charmed
quarks do (or do not) follow the flow as well: for that one should
look at spectra of $D,J/\psi$ and other charmed particles.

\subsection{Qualitative effects of the QCD phase transition on flow}
  We will not have time to discuss details of the
hydrodynamics calculations, which reproduce these data.
 Let me only tell why  RHIC collision  energy 
is so special.

\begin{figure}[h]
\begin{center}
\leavevmode
\vskip -0.5in
\includegraphics[width=2.9in,angle=270]{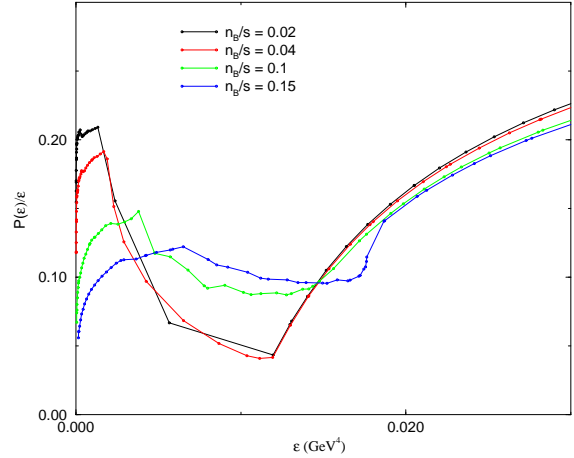}
%\includegraphics[width=2.9in, angle=270]{lifetime.eps}
%\vskip 0.5in
\end{center}
\caption{ \label{fig_soft} The dependence of the
  pressure-to-energy-density
$p/\epsilon$ ratio on the energy density along the adiabatic paths
with different baryon-to-entropy ratio.
RHIC corresponds to the curve with the lowest baryon density,
for which the contrast is the largest.
 }
\end{figure}

The QCD Equation of State (EOS) has  a phase transition, so
 three respective stages of the ``acceleration
history'' can be identified:
(i) the matter is produced an the QGP phase ($e>e_Q$), in which  the 
matter accelerates rapidly because of substantial pressure $p\sim (1/3) 
\epsilon$; 
(ii) then the so called mixed phase follows ($e_{H}<e<e_{Q}$), 
in which the matter is very  soft, $p/\epsilon \ll 1$, and it 
basically free streams with zero acceleration; till finally 
 (iii) the matter completely goes into a hadronic phase ($e<e_H$), 
in which the relativistic pions produces   substantial
 pressure and acceleration again.  
The
effective EOS calculated along the appropriate adiabatic paths
\cite{HS_softest} from resonance gas plus QGP (from lattice)
% shown in Fig.(\ref{}) ?  
is shown in   Fig.\ref{fig_soft}  in the form $p/\epsilon$ versus
$\epsilon$. The effect of the bag pressure on QGP is seen in this
figure 
as a strong dip of this ratio, toward the
so called {\em  softest
  point },  the minimum of $p/\epsilon$.
Since gradient of p is the driving force and $\epsilon$ is the mass to
be moved, the acceleration of matter is proportional to this ratio.
Its small value means that the mixed phase is much softer than both
relativistic pion gas and high-T QGP on both sides of it. 
What this picture shows is that one can  expect more robust
hydro flow when the QGP produced is significantly more dense than
at $T=T_c$, the ``soft QGP''. Heavy ion collisions at SPS energies were not high enough to
have a substantial contribution of the first -- QGP -- stage,
which happens for the first time at RHIC.

%%%%%%%%%%%%%%%%%%%%%%%%%%%%%%%%%%%%%%%%%%%%%%%%%

%%%%%%%%%%%%%%%%%%%%%%%%%%%%%%%%%%%%%%%%%%%%%%%%%%%%%%%%%%%%%%

  Compared to  strong variation of the $p/\epsilon$, 
the observed dependence of the flow  on
the {\it collision energy} 
appears to be  weak. 
While at low BEVALAC/SIS/AGS energies flow velocity $v_t$
steadily grows, for  SPS and RHIC the difference is  not that large. 
However, the relative contribution of stages (i) and (iii)
have exchanged
places, with QGP stage dominating flow values at RHIC.

  Summarizing:  
 Only with
sufficiently high collision energy of RHIC we get to the regime in which
 the transverse expansion at the QGP stage is dominant.
While large QGP pressure  drives the initial
transverse expansion,  the ``dark energy'' (the bag constant
or vacuum pressure)
works against it.

%%%%%%%%%%%%%%%%%%%%%%%%%%% Fig. 7 %%%%%%%%%%%%%%%%%%%%%%%%%%%%%%%%%%%%%%%%%
\begin{figure}[htbp]
\centering
\epsfig{file=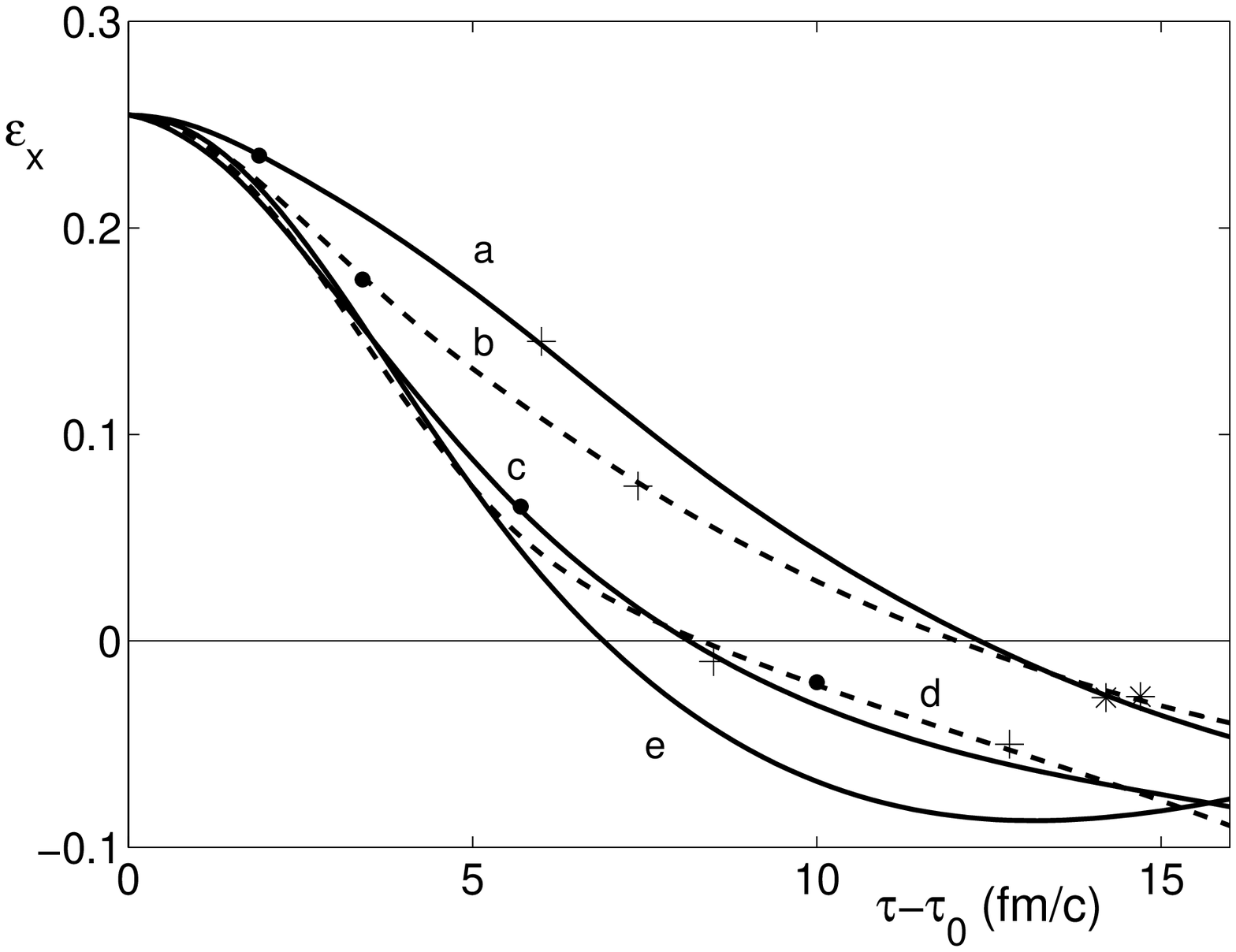,width=7cm}
\epsfig{file=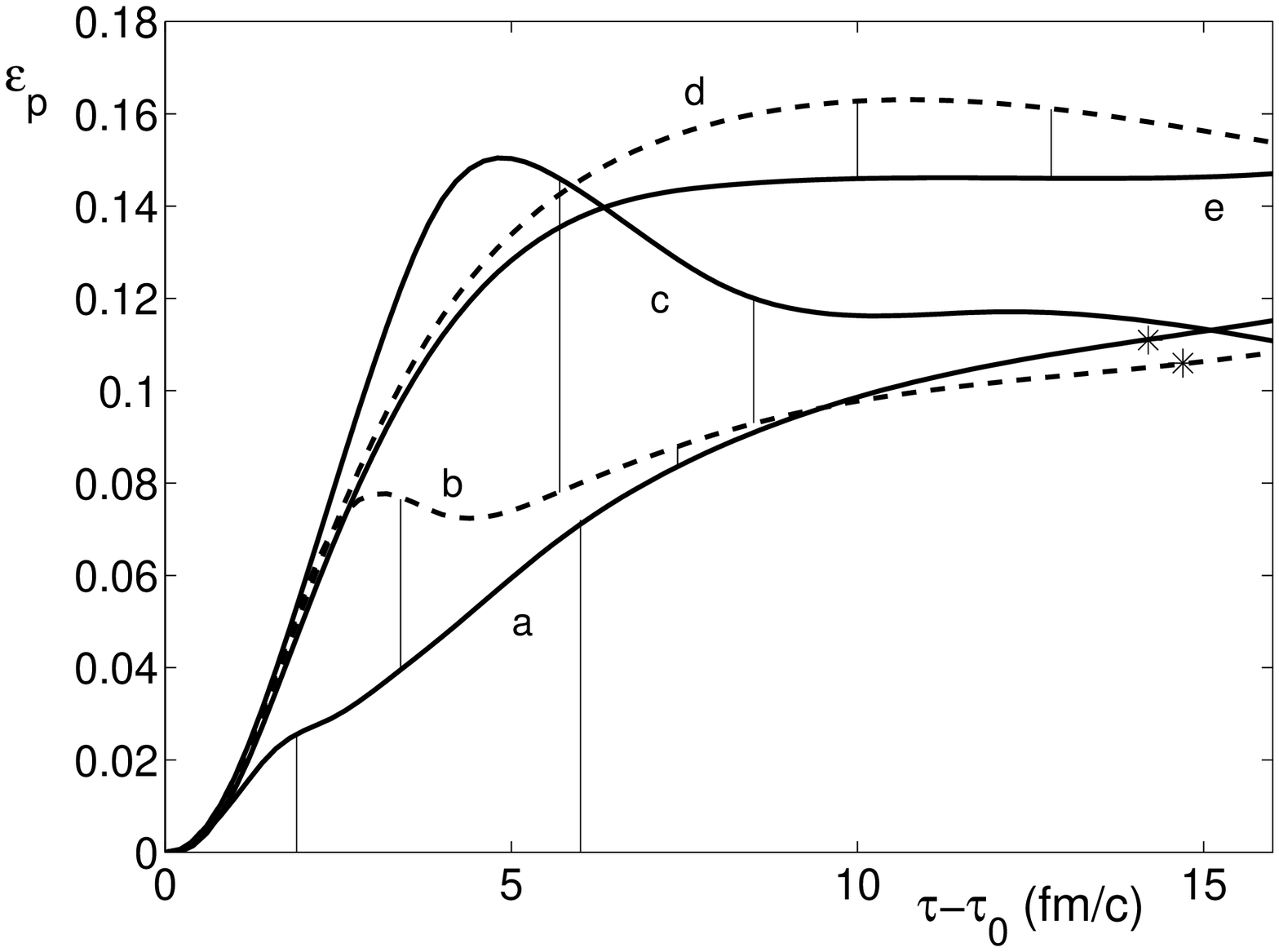,width=7cm}
\caption{Time evolution of the spatial ellipticity $\epsilon_x$, the 
  momentum anisotropy $\epsilon_p$, and the radial flow
  $< v_\perp >$. The labels {\tt a, b, c} and {\tt d} denote 
  systems with initial energy densities of 9, 25, 175 and 25000 
  GeV/fm$^3$, respectively, expanding under the influence of EOS~Q. 
  Curves {\tt e} show the limiting behavior for EOS~I as $e_0\to\infty$ 
  (see text). In the second panel the two vertical lines below each 
  of the curves {\tt a-d} limit the time interval during which the 
  fireball center is in the mixed phase. In the first panel the dots
  (crosses) indicate the time at which the center of the reaction zone 
  passes from the QGP to the mixed phase (from the mixed to the HG
  phase). For curves {\tt a} and {\tt b} the stars indicate the
  freeze-out point.; for curves {\tt c-e} freeze-out happens outside
  the diagram. 
 \label{fig_kolb_v2timing}}
\end{figure}
\vspace*{-0.4cm}
%%%%%%%%%%%%%%%%%%%%%%%%%%%%%%%%%%%%%%%%%%%%%%%%%%%%%%%%%%%%%%%%%%%%%%%%%%%%

\subsection{Elliptic Flow }
\label{sec_Elliptic_Flow}
 
In non-central  heavy ion collisions 
secondary particles emerge with
an nontrivial ``elliptic'' flow pattern. The reason for it
is that while ``spectator'' nucleons
fly down the beam pipe, the produced fireball
has an almond shape.
If pressure develops in the system, its gradient
is larger in the impact parameter direction (the x-direction) 
than in the longer y-direction.
 Then, the
excited matter expands preferentially in the x-direction.
This is quantified by
$v_i$ harmonics defined as\footnote{We will not have time to discuss
odd harmonics like $v_1$, which are nonzero away of mid-rapidity
and measured at different collision energy including RHIC.   }
\be     {dN \over d\phi} = {v_0 \over 2\pi } + {v_2 \over \pi } 
\cos( 2\phi ) +   {v_4 \over \pi }  \cos( 4\phi ) +
                                   \cdots \ee
which are   measured experimentally\footnote{Let me not go into an
issue of the definition of the event plane. With several thousands of
tracks seen in large detectors it is no longer a problem.}.

Each of $v_i$ is a function of centrality (the impact parameter $b$),
rapidity $y$, transverse momentum $p_t$ and, last but not least, the
particle type. All of those have been studied, but we have no
time to go into these details. Just trust me, that hydro  works
nicely, for 99 percents of all secondaries or for $p_t< 2 \, GeV$.
The 
data restrict the underlying EOS, the only theoretical hydro input,
and the best description corresponds to
the latent heat of about $0.8\,\mbox{GeV/fm}^{3}$, which is in good
agreement with the lattice predictions.

The important feature of elliptic flow is {\em self-quenching}.
The time evolution of the  spatial ellipticity $\epsilon_x$, the 
  momentum anisotropy $\epsilon_p$, and the radial flow
  $< v_\perp >$ are shown in Fig.\ref{fig_kolb_v2timing}.
As the system expands, the
eccentricity $\epsilon_2$ decreases.  Since $\epsilon$ 
is the driving force behind the elliptic flow, the  elliptic
flow develops earlier than the radial one.
This is why
 it is so important for understanding the EOS of the QGP.

The next Fig.\ref{fig_v2_growth} makes use of one  important fact:
 centrality dependence
of $v_2$ is basically a response
to 
the initial $spatial$ anisotropy of the system, quantified by the parameter
$$
   \epsilon \equiv \frac{ \langle y^{2}-x^{2}
\rangle } {
   \langle y^{2} + x^{2} \rangle },
$$
and so  plotting $v_2/\epsilon$
one basically eliminates a geometric aspect of the
problem and finds all points at some universal curve, see 
 Fig.\ref{fig_v2_growth}(a).
 
The main message of this figure is 
that $v_2$ at RHIC grows further with particle multiplicity.
 It has been theoretically predicted
already at QM99, before RHIC was even completed. The parts (b,c)
of the figure shows how the $v_2$ magnitude was expected to depend
on collision energy\footnote{Other authors such as Ollitraught
and Heinz et al have used fixed freezeout and
predicted energy-independent $v_2$.}, from Teaney et al \cite{hydro}.

\begin{figure}
\begin{minipage}{7.cm}
\centering
\includegraphics[width=7cm]{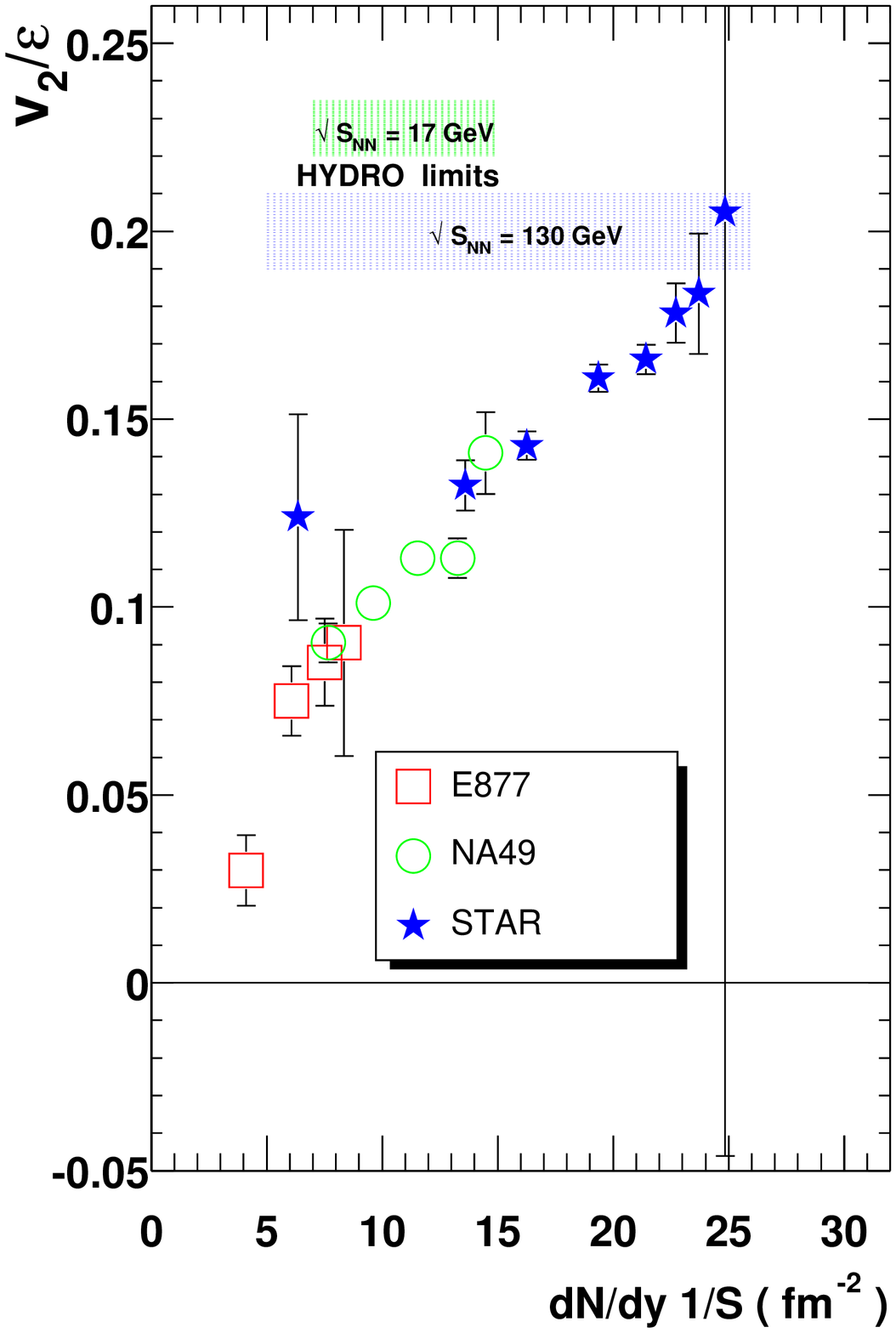}
 \end{minipage} \hspace{1cm}
\begin{minipage}[c]{7.cm}
 \centering 
\includegraphics[width=6.cm]{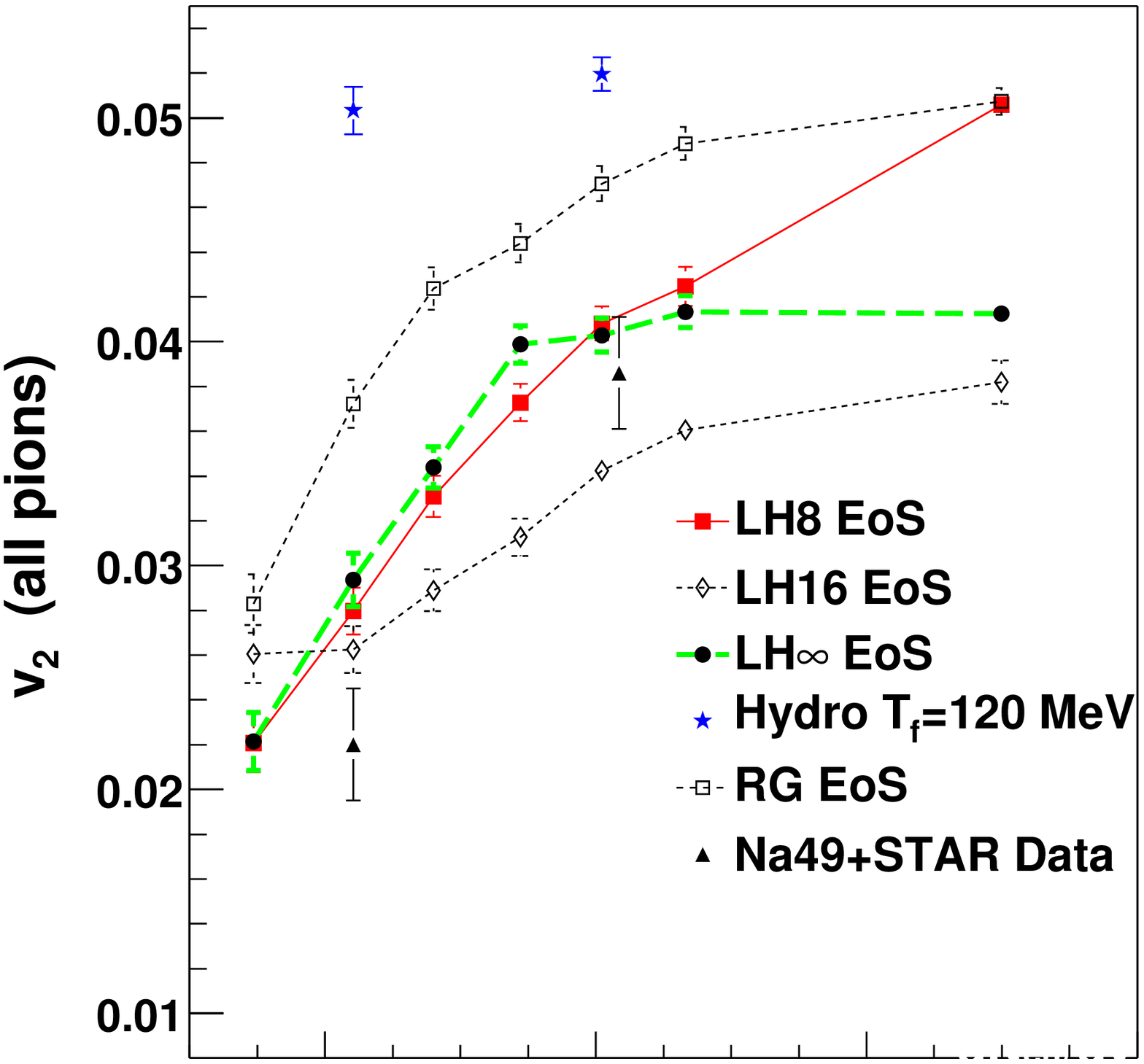}
\includegraphics[width=6.cm]{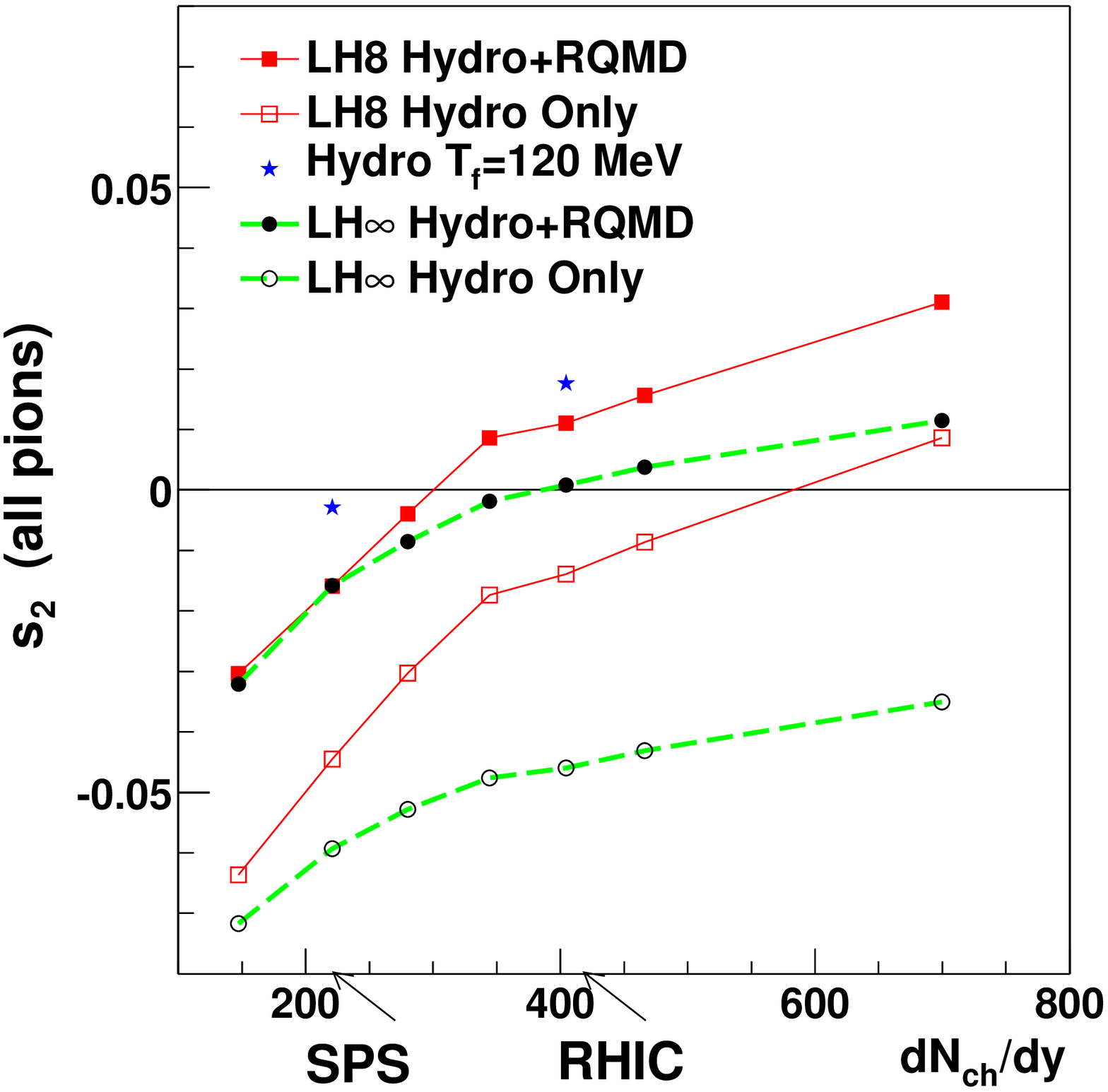} \end{minipage}
\caption{ \label{fig_v2_growth}
(a) The compilation of elliptic flow  (the ratio of $v_2/s_2$)
dependence on collision energy (represented by the particle multiplicity).  
(b,c) Elliptic flow predicted by hydro calculation
by Teaney et al for different EoS. The
 curve with the latent heat (LH) =800 $MeV/fm^3$ is
the closest to the lattice EoS, and it is also the best fit to
  all
flow data at SPS and RHIC}
 \end{figure}

\subsection{The limits to ideal hydro}
\label{sec_viscous_hydro}
  
  It is not easy to reproduce hydro results using transport models.
In fact it is only possible 
 if the quasiparticle rescattering is increased by
big factor relative to the pQCD expectations. 
The following Fig.\ref{fig_GM} from the kinetic studies by
Gyulassy and Molnar \cite{GM} shows how the measured
effect (boxes) can be reached while the
matter opacity (= density times the cross section)
 grows. The smallest
value on the plot roughly corresponds to gg scattering in pQCD,
and one can see that it leads to no elliptic flow (or other
collective effects) whatsoever.

\begin{figure}
\centering
\includegraphics[width=8.cm]{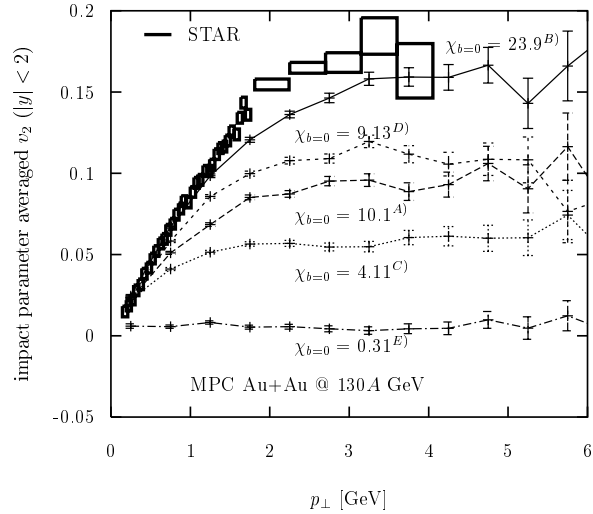}
\caption{
\footnotesize
Impact parameter averaged gluon elliptic flow
as a function of $p_\perp$ for Au+Au at $\sqrt{s}=130A$ GeV
with transport opacities $\chi_{b=0} = 0.31$, 4.11, 9.13, 10.1 and 23.9
for $b=0$.
Identical to the charged hadron elliptic
flow if the gluons are hadronized via local parton-hadron duality.
\label{fig_GM}
}
\end{figure}

The  
     conceptual basis of the hydrodynamics\footnote{Although
  often  hydrodynamics
 is treated as a consequence of kinetic  equations, these two approaches
have a different applicability ranges. Indeed,
 the stronger the interaction 
   the  better hydro works.  The  latter  approach,  on   the
 contrary, was never formulated but  for  weakly  interacting
 systems.
 }
  is  very  simple: it is just
 a set of local conservation laws for the
 stress tensor ($T^{\mu\nu}$) 
and for the conserved currents ($J_{i}^{\mu}$), 
%$\partial_{\mu}T^{\mu \nu}=0$ and $\partial_{\mu}J_{i}^{\mu}=0$.  
\begin{eqnarray}
 \partial_{\mu}T^{\mu \nu}&=& 0   \\ 
 \partial_{\mu}J_{i}^{\mu}&=& 0  \nonumber
\end{eqnarray}
In equilibrium, $T^{\mu\nu}$ and $J_{i}^{\mu}$
are related to the bulk properties of the fluid by
the relations,
%$T^{\mu \nu} = (\epsilon + p) U^{\mu} U^{\nu} - p g^{\mu \nu}$ and 
%$J_{i}^{\mu} = n_{i} U^{\mu} $ 
\begin{eqnarray} \label{eqn_tmunu}
T^{\mu \nu} &=& (\epsilon + p) u^{\mu} u^{\nu} - p g^{\mu \nu}   \\
J_{i}^{\mu} &=& n_{i} u^{\mu} \nonumber
\end{eqnarray}
Here $\epsilon$ is the energy density, $p$ is the pressure,
$n_i$ is the number density of the corresponding current, and 
$u^{\mu}=\gamma(1, \bf{v})$ is the proper velocity of the fluid. 
In strong interactions, the conserved currents are isospin ($J_{I}^{\mu}$),
strangeness ($J_{S}^{\mu}$), and baryon number ($J_{B}^{\mu}$). 
For the hydrodynamic evolution, 
isospin symmetry is assumed and  the net strangeness is set to zero;
 therefore only the baryon current $J_{B}$ is considered below.
These equations are valid if the microscopic length (such as mean free
paths
of constituents $l$) is small compared to the size of the system
considered\footnote{Note that this is the same condition as for other
macroscopic approaches, e.g. thermodynamics: however historically
the hydrodynamical models met much more resistance than say
statistical models. }
$L$. 
Inclusion of dissipative effects, to the first order
in $l/L$, is possible. Of course,
the modified stress tensor and the currents no longer conserve the
entropy,
which is monotonously increased during the evolution. 

     Dissipative corrections to the  stress  tensor  and  the
 current can be written as follows
 \be \delta T_{\mu\nu}=\eta(\nabla_\mu u_\nu + \nabla_\nu u_\mu
 -{2\over 3}\Delta_{\mu\nu}\nabla_\rho u_\rho)+\xi(\Delta_{\mu\nu}\nabla_\rho u_\rho) 
  \ee
                                                        
  \be \delta J_\mu =k ({\eta T \over \epsilon+p})^2 \nabla_\mu (\mu_B/T) \ee
  where the three coefficients  $\eta,\xi,k$         are called the shear
 and  the  bulk  viscosities  and  the   heat   conductivity,
   respectively. In  this  equation  the  following  projection
 operator onto the matter rest frame was used:
  \be \nabla_\mu\equiv\Delta_{\mu\nu}\partial_\nu, \,\,\,
  \Delta_{\mu\nu}\equiv g_{\mu\nu}-u_\mu u_\nu \ee
  
    Two general comments may be made in connection with these
 definitions. First, for matter with vacuum quantum  numbers,
  characterized by  the  temperature  only,  one  cannot  even
  introduce the notion  of  heat  conductivity.  Second,  bulk
 viscosity effects are  very  elusive,  in  particular,  they
 vanish both in ultrarelativistic  and  nonrelativistic  gas.
 So, the main non-equilibrium effect is connected  with  $shear$
 viscosity. Since, as we will show below, the late-time flow
 approaches
%%%%%%%%%%%%%%%%%%%%%%%%%%%%%%%%%%%%%%%%%
radially symmetric Hubble flow without shear, the viscosity effect
has a curious self-quenching property. 
  
It is further useful to relate the magnitude of the
viscosity coefficient $\eta$ to a more physical observable. As such
one can use the sound attenuation length.
\index{viscosity: sound attenuation length}
 If a sound wave have frequency
$\omega$ and the wave vector \q, its dispersion law (the pole position) is
\be \omega=c_s q-{i\over 2} \q^2 \Gamma_s, \qquad \Gamma_s\equiv {4\over 3}
{\eta \over \epsilon+p}\ee

 In order to show how such parameter appears in hydro equations, let
 me give an example of
 the 1+1d  boost-invariant  solution
  \cite{Bjorken}, corresponding to 
 the velocity
$ u_\mu=(t,0,0,z)/\tau $
where $\tau^2=t^2-z^2$ is the proper time. In
     this 1-d-Hubble regime 
  there   is   no   longitudinal
 acceleration at all: all  volume  elements  are
 expanded linearly with time  and  move  along   straight
 lines from the collision point. 
 Exactly as in the Big Bang, for each "observer" ( the volume
 element ) the picture is just the same,  with  the  pressure
 from the left  compensated  by  that  from  the  right.  The
 history is also the same for all volume
elements, if  it  is  expressed  in  its  own
 proper time $\tau$.
Thus the
 entropy conservation  
becomes the following (ordinary) differential equation in proper time $\tau$
\be {ds(\tau) \over d\tau}+{s\over \tau}=0 \ee
which has the obvious solution 
$
 s={const \over \tau}$.

 Including
   first dissipative terms into our equations one has instead
  \be 
{1 \over\epsilon+p} {d\epsilon \over d\tau} ={1\over s}{ds\over d\tau}=-{1 \over
   \tau}\left(1-{(4/3)\eta+\xi \over (\epsilon+p)  \tau}\right)
      \ee
Note that ignoring $\xi$ one finds in the r.h.s. exactly
the combination which also appears in the sound attenuation, so the correction
to ideal case is  $(1-\Gamma_s/\tau)$. Thus the length $\Gamma_s$
   directly tells us the magnitude of the dissipative corrections. 
Since the correction is negative,
 it reduces the rate of the entropy decrease with time.
Another way to say that, is that the total positive sign shows that
some amount of entropy is
generated.

In order to understand qualitatively what viscous term would do it is 
convenient to use the simplest analytic example, Bjorken boost
independent
flow, and calculate what is added to the ideal isotropic pressure
terms
in the stress tensor. The result is that in longitudinal $z$ and
transverse
directions it is very different
\be p_z\rightarrow p-{4\over 3}{\eta\over\tau} \qquad
p_\perp\rightarrow p+{2\over 3}{\eta\over \tau}\ee
Viscosity works against the longitudinal pressure, this is
understandable:
but why it helps the transverse one? The reason is particle
distribution over momenta is deformed, in longitudinal direction
it has less particles with large $p_z$ than the thermal one because
those ``jump'' to other cells. But then distribution over $p_\perp$
should
be somewhat wider, to keep mean chaotic energy the same. (Note that
the sum of all 3 pressures is unchanged.)

  Let us now discuss what is the value of QGP  viscosity, following
Teaney \cite{Teaney_visc}, who
 shown that one can also determine
the viscous correction to particle distribution
\begin{eqnarray}
\label{correction}
   f = f_{o}(1 + \frac{\eta}{s T^3} p^{\alpha}p^{\beta} 
   \left\langle \nabla_{\alpha}u_{\beta} \right\rangle)\;.
\end{eqnarray}   
where the gas is assumed to be a Boltzmann gas (otherwise $f_{o}(1 \pm
f_{o})$ is needed for Bose/Fermi ones). 
The tensorial structure makes is clear that
any deviations are expected to be quadratic
in momenta.

%%%%%%%%%%%%%%%%%%%%%%%%%%%%%%%%%%%%%%%%%%%%%%%%%  
\begin{figure}[h]
%\begin{minipage}[c]{6.cm}
 \centering 
\includegraphics[width=8.cm]{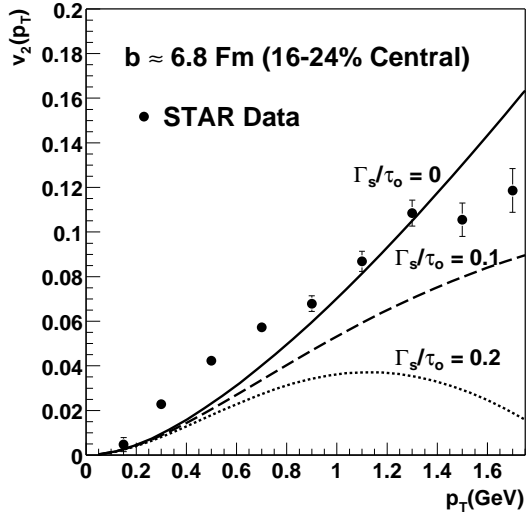}
% \end{minipage}
%\begin{minipage}[c]{6.cm}
% \centering 
%\vskip .3cm
%\includegraphics[width=6cm]{mCorrections.eps}
%\end{minipage}
\caption{ Elliptic flow $v_2$ as a function of $p_T$ for different 
values of $\Gamma_s/\tau_{o}$. The data points 
are four particle cumulants data from the STAR collaboration.
Only statistical errors are shown.
%(b)  Viscous correction $\delta R^2$  
%for HBT radii $R_O$, $R_S$, and $R_L$
%relative to ideal blast wave predictions $(R^2)^{(0)}$ .
\label{fig_visc_v2_radii}
}
%\end{minipage}
\end{figure}

 Such deviations  from hydro predictions are indeed seen
in spectra. 
In  Fig.\ref{fig_visc_v2_radii} we show it for 
the  elliptic flow
parameter $v_2$. Since its value is determined
at sufficiently early times
 -- about 3 fm/c -- the deviation should
 correspond to  the QGP  phase.
The results for different $\Gamma_s/\tau$  shown in 
Fig.\ref{fig_visc_v2_radii}
 deviate from ideal hydro curve 
at $p_\perp\approx 1.6 \, GeV$ which indicates $\Gamma_s/\tau\sim
0.05$ or so. Substituting here the relevant time $\tau\sim 3 fm/c$ we get
 $\Gamma_s\sim .15 fm$. Strong coupling result for typical $T\sim 200
 \, MeV$ at the time gives $\Gamma_s\sim 0.1 fm$, while weak coupling
one would predict much larger value  $\Gamma_s\sim 2 fm$ or so.   
About the same value appears from a gluon cascade with the enhance cross
section by Molnar and Gyulassy \cite{GM} we mentioned above.

\subsection{  The ``Little Bang'' versus the Big Bang  }

 Completing the chapter on heavy ion collisions, let me make
 a more general comments
 emphasizing multiple amusing analogies
between the heavy ion physics
 and the Big Bang
cosmology.  

 First of all, the same
hot/dense
hadronic matter, which heavy ion physics is trying to produce 
in the laboratory, has  been present at an appropriate time
(microseconds) after Bing Bang.
More specifically, in both cases 
the matter goes through the QCD phase transition, from the
 {\em Quark-Gluon Plasma}
we discussed in the preceding chapter to the hadronic phase.
On the phase diagram 
the Big Bang path proceeds along the T axis downward , since the
baryonic density in it is tiny, $n_B/T^3< 10^{-9}$.
For heavy ion collisions  the baryonic charge is
small but still important.
experiments.

The fireball created in the heavy ion collision of course explodes, as
its high pressure cannot be contained: this is the {\em Little Bang}
we referred to in the title of this section.
The {\em Big Bang} is a cosmological explosion, which proceed against
the pull of gravity. So, 
the Little Bang is a laboratory simulation of its particular stage.
So,
 the first obvious similarities between
the ``Little Bangs" 
and cosmological ``Big Bang" is that both
are violent explosions. 

In both cases the entropy is mysteriously produced at some early stage,
and is approximately conserved later.

Expansion of the created
hadronic fireball  approximately follows the same Hubble law
as its bigger relative,
$v(r)=H r$ with $H(t)$ being some time-dependent parameter,
  although the Little Bang has rather anisotropic (tensorial) H.
However, by the end of the expansion the anisotropy
is nearly absent and local expansion at freezeout at RHIC
is nearly Hubble-like.

 The $final$  velocities of collective motion in the Little Bang are
measured in spectra of secondaries: the transverse velocities
 now are believed to be reasonably
well known, and are not small, reaching about 0.7c at RHIC. 
For the Big Bang there is no end and it proceeds till today:
the current value of the Hubble constant,
$H(t_{now})$ is, after significant controversies 
for years,  believed to be reasonably
well measured.

However already the next important
question one would obviously ask, {\em how exactly
such  expansion rates
are achieved} 
remains
  a matter of hot debates in both cases.
 The observed velocity
of matter expansion at freezeout of the collisions or in Big Bang today
is determined by the earlier acceleration, or
 the Equation of State
(EoS)
of the  matter. 

For the Big Bang the EoS include gravity of 
all matter forms, including the still
mysterious {\em dark matter}, as well as
  the (really shocking) {\em dark energy}
(or the
 {\em
cosmological constant}). Recent experimental observations  
of Supernovae in very distant
Galaxies have concluded  that  the Bing Bang is
$accelerating$
at this time due to it.

Amusingly,   the Little Bang has to deal with the
cosmological
constant as well. Of course,  gravity is not important here,
but going from QGP to hadronic matter one has to think about
a definition of the zero energy density and pressure, or the
so called bag constant between the two phases. As we will see,
the cosmological term exists in this case, but it
has the opposite sign and tries to
$decelerate$ expansion of QGP.
The EoS 
 is thus effectively very soft near the QCD phase transition: respectively
the magnitude
of the flow observed at AGS/SPS energies is not that large. With RHIC
we had the first chance to go well beyond into the QGP domain,
with harder EoS $p\approx \epsilon/3$ and more robust expansion.

 In a Little Bangs  
the  observed hadrons
(like microwave  cosmic photons) are seen at the moment of their last
interaction, or as we call it technical, at
 their {\it freeze-out stage}.
 In order to look deeper, one uses rare hadrons
particles with smaller cross sections, such as $\Omega^-$ hyperons,
which decouple earlier, or even {\em penetrating probes}
\cite{Shu_QGP} 
leptons or photons
which penetrate through the whole system. We return to their
discussion
in the next chapter.

The next comparison I would like to make here deals with
the issue of fluctuations.  Very impressive
 measurements of the microwave background anisotropy  made
in the last decade have taught us a lot about cosmological
parameters. First
the dipole component was found -- motion of the Solar system
relative to the microwave heat bath, and then a
very small  ($\delta T/T\sim 10^{-5}$) chaotic
fluctuations of T originated from plasma oscillations  
at the photon freeze-out. It has been possible lately
to measure some interesting structures in fluctuations of cosmic
microwave background, with angular
momenta
$l\sim 200$. 
The theoretical predictions for these were available
since it is related to primordial plasma-to-gas transition at
temperature $T\sim
1/3 eV$.
Primordial fluctuations of sufficiently long wavelength are attenuated 
by the gravitational instability, till they reach the stabilization moment
$\tau_{stabilization}$
when the instability is changed to a regime called
{\em Sakharov acoustic oscillations}. Hydrodynamics tells us that fluctuations
disperse during this period with the (current) speed of sound, reaching the
so called  {\em sound horizon} scale 
\be \label{eq_sound_hor}
r_{s.h.}=\int_{\tau_{stabilization}}^{\tau_{observation}} c_s(t) dt\ee
This scale is the physical size of spots observed, and correspond to 
a peak at $l\approx 220$ recently observed.

In heavy ion collisions  similar studies are in its
infancy. There exist of course the {\em global ellipticity} of the
event,
related to elliptic flow and nonzero
impact parameters:  we will discuss it below in detail in
section \ref{sec_Elliptic_Flow}. 
 We do not yet see reliable
signals for either mean values of higher harmonics, or
their fluctuations.
Nevertheless, since in the mixed phase the transition of high density
QGP phase into low density hadronic gas
also should be characterized by some instabilities,
the sound horizon is also a useful concept to define the
observed spectrum of the fluctuations\footnote{The paper on that
  subject I am working on is in progress at this point.}
Although we do not yet have
any observations of higher harmonics and their fluctuations,
 I think there is a chance to see
 eventually ``frozen plasma oscillations'' in this case as
 well. After all, RHIC produced millions of events, while 
The Big Bang remains the  only one! At the end of the day, the
amount of information taken about the Little and Big Bangs are
comparable\footnote{In the final analysis, both
are determined by number of pixels in the
detectors and computer storage.}.

\section{Transport properties of the QGP}

The perturbative approach to transport
 is well summarized by P.Arnold et al
\cite{AMY}.
The high temperature shear viscosity
in a gauge theory with a simple gauge group
(either Abelian or non-Abelian)
has the leading-log form
\begin {equation} \label{eq_visc_pert}
    \eta = \kappa \, {T^3 \over g^4 \, \ln g^{-1}} \,,
\end {equation}
where $g$ is the gauge coupling, which is presumed to be small.
For the case of $SU(3)$ gauge theory 
 the leading-log shear viscosity coefficient $\eta$
for various numbers of light $m<<T$ fermion species
are shown in the Table \ref {shear_table}. One may conclude from it
that $\eta/s $ should be very large compared to 1.

\begin{table}[tb]
\tabcolsep 10pt
\begin {center}
\begin{tabular}{cc}
$ \quad \nf \quad $ & $ \eta \times (g^4 / T^3) \ln g^{-1} $ \\ \hline 
0 & \phantom{1}27.126  \\
1 & \phantom{1}60.808  \\
2 & \phantom{1}86.473  \\
3 & 106.664  \\
4 & 122.958  \\
5 & 136.380  \\
6 & 147.627
\end{tabular}
\end {center}
\caption
    {%
    \label{shear_table}
    Leading-log shear viscosity as a function of the
    number of (fundamental representation) fermion flavors with $m \ll T$,
    for gauge group $SU(3)$.
    }
\end{table}

 We will discuss the  $\cal{N}$ = 4  
supersymmetric Yang-Mills theory below, but let me jump ahead and
mention here the
strong coupling result for the viscosity
 by Policastro, Son and Starinets \cite{PSS}.
Their main result for shear viscosity is 
\be \eta={\pi \over 8} N_c^2 T^3\ee
which, after it is combined with the results for
thermodynamics \cite{thermo}, shows that the 
dimensionalless viscosity-to-entropy ratio is small, specifically
\be {\eta\over s}={1\over 4\pi}\ee
They also found other evidences that hydrodynamics is a good theory
in the strong coupling limit.

In subsequent works \cite{Starinetsetal}  more examples of Maldacena-like
correspondence were studied,
and this particular ratio was found to be the same for all cases.
These authors conjectured that it is in fact the lowest possible
value of this ratio, representing  the {\em most perfect}
liquid possible.

\subsection{Rethinking the QGP at RHIC, $T_c<T<2T_c$ }
The earliest suggested QGP signal was
a disappearance of familiar hadronic peaks -- $\rho,\omega,\phi$ mesons --
in the dilepton spectra \cite{Shu_QGP}. Moreover, even small-size deeply-bound
$\bar c c$ states,  $\eta_c,J/\psi$, were expected to melt at
$T\approx T_c$ \cite{MS,KMS}. 
 
\begin{figure}
\begin{minipage}{7cm}
\centering
\includegraphics[width=6.cm]{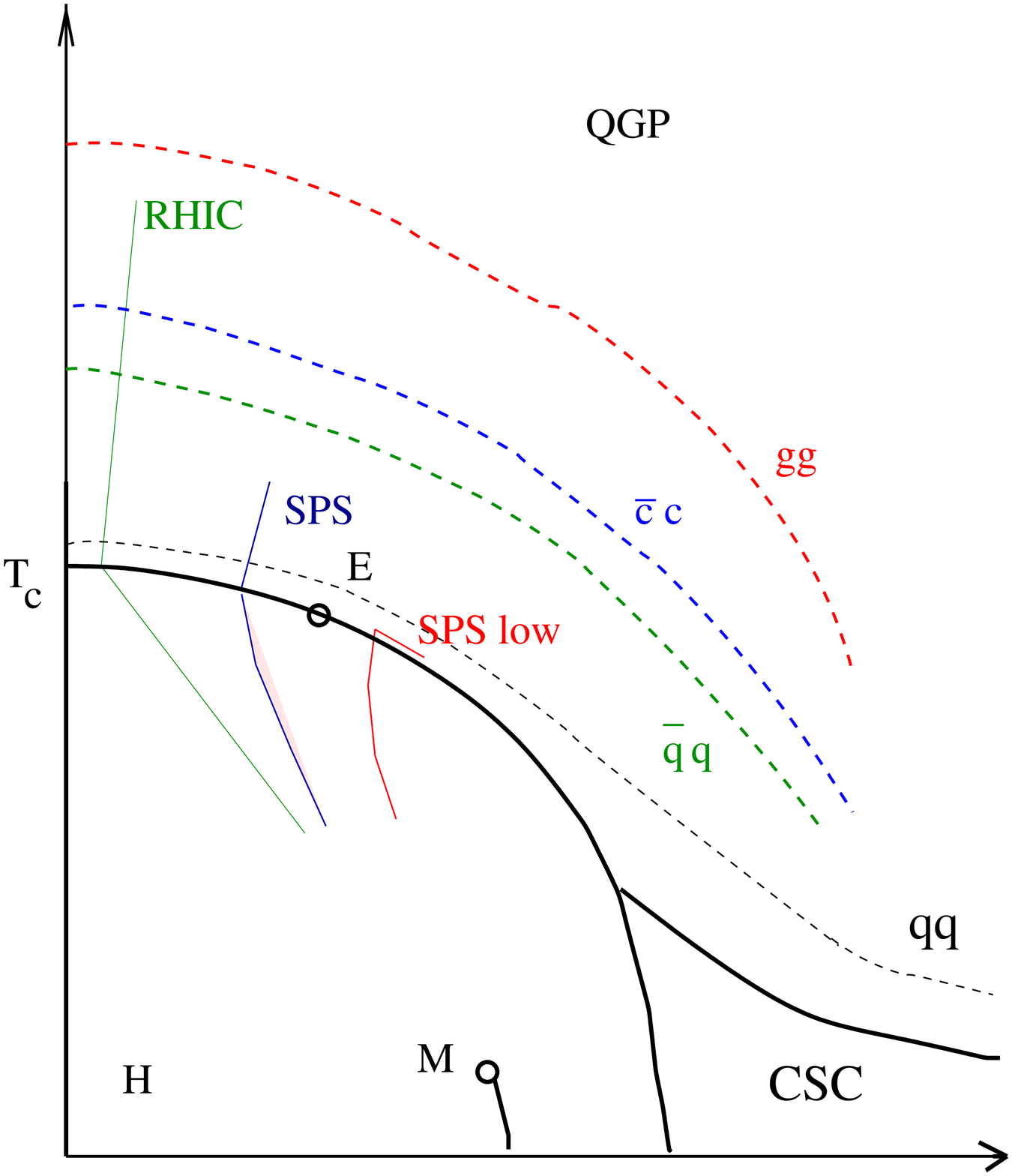}
\end{minipage}\begin{minipage}{9cm}
\vskip .7cm
\includegraphics[width=9.cm]{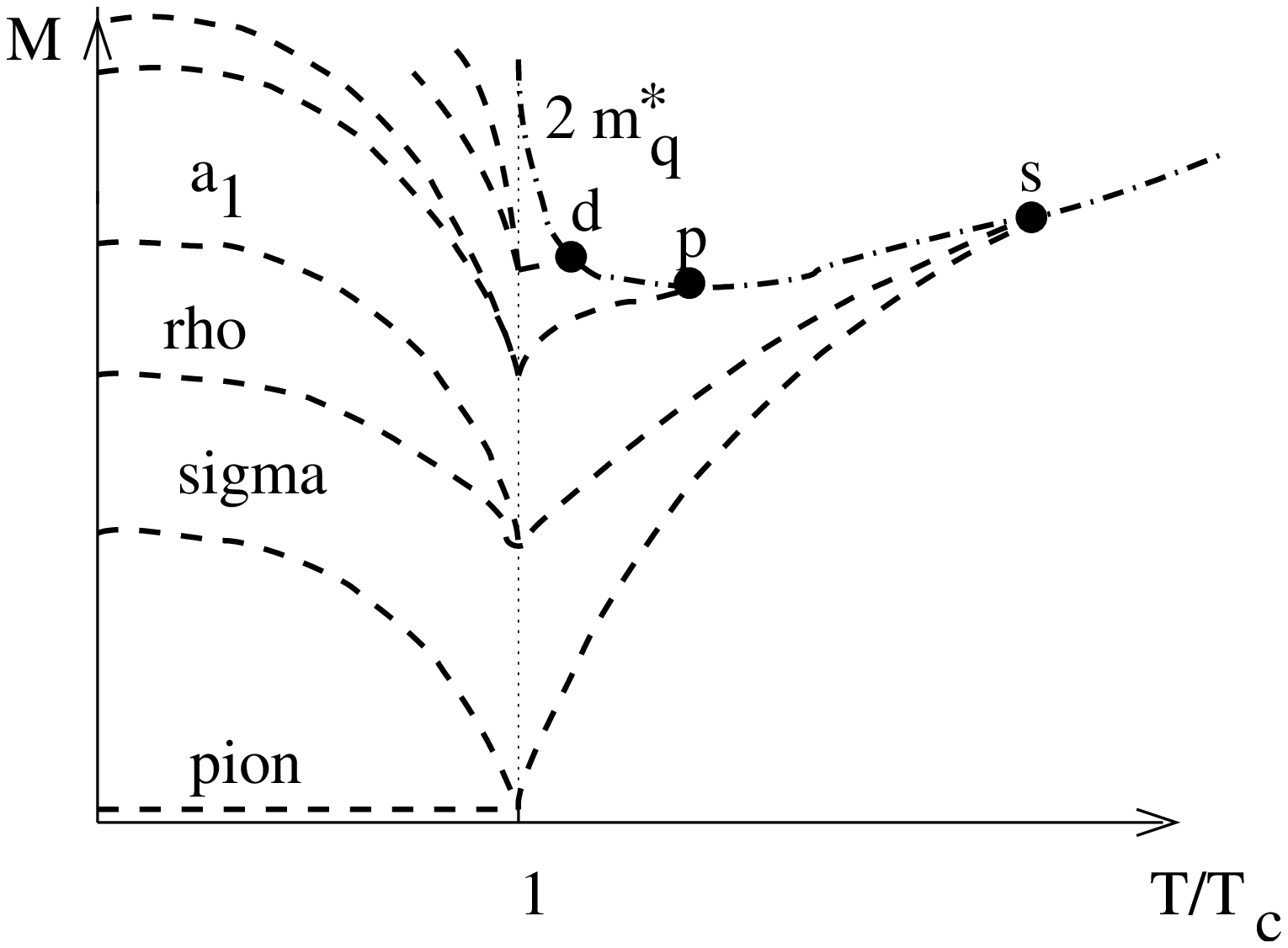}
\end{minipage}
   \caption{\label{fig_masses_T}
Schematic position of several  zero binding lines on the 
QCD phase diagram (a) and of specific
hadronic masses on temperature $T$ (b). In the latter 
the dash-dotted line shows twice the (chiral) effective mass
of a quark. Black dots marked $s,p,d$ correspond to the points where
the binding vanishes for
states with orbital momentum $l=0,1,2...$.
}
\end{figure}

   However, it turns out that
in wide region  $T<4 T_c$ the interaction
is strong enough, causing 
%perturbation theory
%to fail in general\cite{com1}.
% In this letter we 
%will discuss
 bound states of quasiparticles. More
specifically, those are limited by
the {\em zero binding lines} on the phase diagram
 (see Fig.\ref{fig_masses_T}(a)), introduced by I.Zahed and myself
in \cite{SZ_newqgp}.

  Discussion of particular
 hadronic states in QGP have been done many times before, see e.g. 
application of the
sum rules or 
 instantons.
There are also  important  lattice results. It has been shown
\cite{charmonium} that
(time-direction) correlators at $T>T_c$ display significant
deviations
from free behavior, 
in quantitative agreement with predictions in \cite{Schafer:1995df}.
 Their
analysis (by the minimal entropy
method) have suggested existence of light quark resonances
above $T_c$. Most definite are
recent studies \cite{charmonium} which  found (contrary to
earlier expectations) that the lowest 
charmonium states remain  bound  up to $T<T_{\bar c c} \approx (1.6-2) T_c$. 

The bound states of $\bar q q$ can only be colorless mesons
(the octet channel  is repulsive), but in QGP there can
be $colored$ bound states. Quite famous are quark Cooper pairs $qq$
which drive the color superconductivity at sufficiently high density
and low $T$: but pairs themselves should exist outside this region as well.
Gluons can form a number of states with attraction, and there can
also be $gq$ hybrids. A generic reason why we think all of them exist
is that at $T$ close to $T_c$ all quasiparticles are very heavy.

%%%%%%%%%%%%%%%%%%%%%%%%%%%%%%%%%%%%%%%%%%%%%%%%%%%%%%%%

%%%%%%%%%%%%%%%%%%%%%%%%%%%%%%%%%%%%%%%%%%%
Starting with quarks, let us briefly discuss their quantum numbers.
Ignoring current quark masses and instanton effects, we note
than in QGP  the $chiralities$
(L,R) are good quantum numbers. Furthermore,
there are two different modes of
quarks depending on whether $helicity$ is the same as chirality
or not (these states are called ``particle'' and ``plasmino'', respectively). 
 All in all there are $4 N_f^2$ states, connected by continuity
to the pseudoscalar, 
scalar, vector and axial vector nonets in the vacuum 
(see Fig.\ref{fig_masses_T}(b)). 

 Gluons in QGP  have  not 2 but 3 polarizations, with a
 longitudinal  ``plasmon'' mode.
Furthermore, for $gg$ one can have not only 
color singlet (a continuation of the vacuum glueballs)
but also octet binary composites.
 The exotic states $qq,qg$ or octet $gg$ have smaller Casimirs
and need stronger couplings, so they are melted first.

{\bf Our main idea} is that a sequence of loosely bound 
 states (indicated by black dots in  Fig.\ref{fig_masses_T}(b)) is
 of crucial importance for quasiparticle rescattering,
since there its cross section is strongly enhanced. Huge cross sections
induced by low-lying
resonances are well known throughout all parts of physics.
%e.g.  cold neutron rescattering on appropriate isotopes with a level close 
%to zero energy. 
The Breit-Wigner cross section (modulo the obvious spin factors
depending on the channel) is 
\be \sigma(k)\sim  {4\pi \over k^2} {\Gamma_i^2/4 \over (E-E_r)^2+\Gamma_t^2/4}
\ee
For  $E-E_r\approx 0$ the in- and total
widths  approximately cancel: the resulting  
``unitarity limited'' scattering is
 determined by the quasiparticle wavelengths 
which can be very large.
We conjecture that 
this phenomenon significantly contribute to viscosity reduction.

%%%%%%%%%%%%% details on binding from here %%%%%%%%%%%%%%%%%%%%% 

{\bf The issue of $\bar c c$ binding}  can be addressed
 using
the usual non-relativistic 
Schrodinger equation, which for 
 standard radial wave function
$\chi(r)=\psi/r$   has the usual form,
%\be {d^2 \chi \over dr^2}+2m(E-V_{\rm eff})\,\chi=0\ee 
with the reduced mass $m=m_c/2$ and an effective potential
including T-induced screening of the charge \cite{Shu_QGP},
which we will write in a traditional
 Debye form 
$ V=-(4\alpha_s(r)/3r) exp(-M_D r).$
 We have chosen the ($T$-dependent) inverse screening mass    
to be our  unit of length,  so that
 the equation to be solved reads 
\be {d^2 \chi \over dx^2}+\left(\kappa^2+{4m_c\over 3M_D} {\alpha_s(x)\over
  x}e^{-x} \right)\chi=0\ee
with $\kappa^2=m_c E/M_D^2$. Furthermore,
at a zero binding point $\kappa=0$. So, if
the coupling constant $\alpha_s$ would  not run and be just a
constant, all pertinent parameters 
appear in a single combination. Solving the equation, one  finds
 the zero binding condition to be 
\be \label{eqn_comb}
  {4m_c\over 3M_D} \alpha_s = 1.68 \label{cond}\ee
For example,  using $4/3\alpha_s=0.471$
 and $m_c=1.32 \, {\rm GeV}$, as
 Karsch et al \cite{KMS} did long ago,
 one finds a restriction on the screening  mass
$M_D< M_D^{crit}=0.37\, {\rm GeV}$. 
Lattice measurements of the screening masses for the near-critical QGP 
\cite{Karsch_screening}  found that (not very close to $T_c$)  
 $M_D/T\approx (2.25\pm .25)$ in a relevant  range of $T$. 
If so,  the condition
(\ref{cond}) is satisfied marginally if at all, and these authors 
concluded that charmonium
s-wave states $\eta_c,J/\psi$ cannot exist inside the QGP phase.

A loophole in this  argument is the assumption that
the gauge coupling is a
constant, with the {\it same} value as 
 in the in-vacuum charmonium potential. 
However in-vacuum potential includes large confining linear term,
absent at $T>T_c$.  
  The true form of $V_{eff}(r)$ remains unknown.  At $T>T_c$  
 nothing prevents  the QCD coupling from running
at larger distances to 
larger values 
 until it is stopped at the screening
mass scale.
 Very close to the  
critical point, where the screening mass is very small,  
the coupling can reach  $\alpha_s\sim 1$ and presumably be limited by some
nonperturbative phenomena. 

we use a simple model-dependent potential, in which 
 the charge  continues to run till it is limited 
by $\alpha_s<1$.
%In Fig.\ref{fig_g_binding}(a) we compare  such potential at several
%values of $T$ to its
% version with constant coupling.
% The cusp
%in the potential reflects on that, and it occurs at about the
%screening length $r\,M_D\approx 1$.
%
Such potential with the running coupling does indeed provide a more liberal 
condition for  charmonium binding, which is  $M_D<0.62\, {\rm GeV}$. This
translates into charmonium zero binding point at 
\be T<T_{\bar c c} \approx 1.6\, T_c. \ee
which  agrees well 
with recent lattice measurements \cite{charmonium}:
we thus conclude that 
our model-dependent  potential has passed its first test.

{\bf The binding of light $\bar q q$ states.} 
 Chiral symmetry for massless quarks
 excludes the usual mass from being developed, and
$L,R$-handed quarks propagate independently. Nevertheless, 
propagating quasiparticles if the QGP have dispersion curves with the
nonzero  ``chiral'' or ``thermal'' mass, 
defined as the energy of the mode at zero momentum $M_q=\omega(\vec
 p=0)$. Perturbatively it is $M_q=gT/\sqrt{6}$ to the lowest order,
the same for both  fermionic modes: {\bf i.} with the {\em same
chirality and helicity} the dispersion curve at small $p$ is
$\omega=M_q+p/3+p^2/3M_q+...$; {\bf ii.} with the opposite chirality and
helicity the mode is often called a  ``plasmino'', its dispersion curve
has a shallow minimum at $p=0.17gT$ with the energy $E_{\rm min}=0.38\,gT$
slightly below $M_q$. 
For a general analysis of these modes see \cite{Wel_99}.

 Lattice data on the quasiparticle
dispersion curves  are rather sketchy, obtained
in a Coulomb gauge~\cite{latt_quasipart_masses}. They can be described by
 $\omega^2=p^2+M^2$, with the following values (at $T=1.5 T_c$) 
\be  \label{eqn_masses}
{m_q \over T}= 3.9\pm 0.2 \hspace{1cm} {m_g \over T}= 3.4\pm 0.3\ee
Note that at such $T$ (which is not very close to $T_c$)
the quark and gluon masses are still quite large, while
 their ratio  is 
very different from the weak coupling prediction $1/\sqrt{6}$. 
At $T\sim 3T_c$ and higher they are somewhat reduced toward
the perturbative values, which however are not reached till presumably
very high $T$.

%If the masses of quasiparticles are large, there is no
%principal difference with the charmonium case. Furthermore,
The effective equation of motion suitable for discussion
of the bound state problem can be obtained by
standard substitution of the covariant
derivatives in the place of momentum and frequency, 
provided the dispersion law is known.
So if
the  dispersion curve can be parameterized as
$\omega=M_q+p^2/2M'+..$,  it has the form of
the nonrelativistic Schrodinger equation, in general with two different
constants $M,M'$. Both for weak coupling and lattice data,
such approximation seem to be accurate    
withing several percents.

Addressing the issue of binding,
we first note that if all effective masses  grow linearly with $T$,
including the screening mass, the explicit $T$ dependence drops out
of (\ref{eqn_comb}) to the exception of the logarithmic
dependence in $T$ left out in $\alpha_s\sim 1/{\rm ln}(T/\Lambda_{QCD})$.
This is why the region of ``strongly coupled QGP'' turns out to be 
relatively substantial. For a qualitative estimate, let us
set the  coupling to its maximum, $\alpha_s=1$. The combination of 
constants is $(4\times 3.9\,T)/(3\times 2.25\,T)=2.3$, larger than the
critical value (\ref{eqn_comb}), so one should expect the 
occurrence of (strong) Coulomb bound states. Although the (plasmon)
gluon modes are somewhat lighter than quarks in (\ref{eqn_masses}),
their Coulomb interaction has a larger coefficient due to
a different Casimir operator for the adjoint representation, 3
instead of 4/3. As a result the effective combination in the potential is
$3 m_g \alpha_s/M_D$, which  is about twice larger than for quarks, and
thus the gluons are bound even stronger (modulo collisional broadening).

Solving the equations one can make a quantitative analysis, using the same
potential as above. We found that the highest temperature $T$ at which
light quark states are Coulomb bound is somehow lower than that of charmonium,
\be
T_{\bar q q}\approx 1.45\, T_c \approx 250 \, {\rm MeV}\,\,,\ee while the 
s-wave $gg$ gluonium states 
remain bound
till higher temperatures  
$
T_{gg}\approx 4 \,T_c $ used in this paper as the upper limit on the
QGP with bound states.
%As an example,  in Fig.~\ref{fig_g_binding}(b) we show the 
%binding energy of two gluons. (In absolute units it reaches 
% about 100 MeV, while it is much smaller MeV for quarks.)
%The  wave functions of all loosely bound
% states are similar to 
% that of a deuteron, with the usual $\chi\sim e^{-\kappa r}$
%behavior outside  the potential.

  How reliable is our approach, based on screened effective potential?
Obviously even the best
  in-matter potential does not include
  all  many-body
effects. However one may think that large size of
states at near-zero binding provides additional stability,
 averaging out  local perturbations.
This is known to be true for large-size Cooper pairs in superconductors,
or ``excitons''  in semiconductors and insulators.
Depending on a number of
parameters, including the density of excitons and temperature, 
this system exhibits various phases, ranging from an ideal gas of excitons to
a liquid or plasma, or even a Bose-condensed gas. On its way from a gas
to a liquid, clustering with 3- and 4-body states play an important role.
Although one cannot directly relate these two problems
(quarks and gluons have $N_c$ and $(N_c^2-1)$ colors respectively, while particles
and holes have simply charges $\pm e$), one may think that in the QGP at $T\sim
T_c$ some of these phenomena may well be there. 

%%%%%%%%%%%%%%%%%%%%%%%%%%
 In a more recent paper, by Brown et al \cite{BLRS}, 
the fate of the  $\bar q q$ bound states
is traced to $T\approx T_c$, where the
Nambu-Goldstone and Wigner-Weyl modes meet.
 the binding of these states is accomplished by
the combination of (i) the color Coulomb interaction,
(ii) the relativistic effects, % including spin-spin forces
and (iii) the quasi-local
 interaction induced by the instanton-anti-instanton
molecules. The spin-spin forces turned out to be small.

 While near
$T_{zb}$ all mesons are large-size nonrelativistic objects bound
by Coulomb attraction, near $T_c$ they get much more tightly
bound, with  the $\sigma$ and $\pi$ 
masses approach zero (in the chiral limit).
%$T_c$.
The wave function at the origin grows strongly with binding, and
the near-local four-Fermi interactions induced by the instanton
molecules play an increasingly more important role as the
temperature moves downward toward $T_c$.

\section{The  \N=4 supersymmetric gauge field
  theory at strong coupling }

\subsection{The main results obtained via the Maldacena duality }

In QCD, as soon as the lowest  states $\pi,\sigma$ hit zero
mass, there is an instability leading to a phase
transition into hadronic (confining
and chirally asymmetric) state. This is however impossible for 
 conformal gauge theories (CFTs), such as  \N=4 supersymmetric
 gauge
 theory. In CFT 
the gauge coupling is allowed to become {\em supercritical} or even large 
 $\lambda\equiv g^2 N_c \gg 1$. 

In order to compare it properly with the QCD
 we remind
that on top of the $SU(N_c)$ gauge fields this theory has 4 gluino-like
fermions 
and 6 scalars\footnote{This is needed by supersymmetries,
to balance the number of degrees
  of freedom $(N_c^2-1)*(2-8+6)=0$, where we use minus for fermions.} .
For 3 colors this is 16 gluons, plus
 64 fermionic d.o.f. (to be compared to 12$N_f$
with anti-fermions in
QCD)
plus 48 scalars.

Further modification of the background metrics \cite{thermo} with a Schwartzschild (black
hole)  has allowed
to 
consider \N=4 theory at finite temperature $T$. 
The metrics used in all calculations to be referred below is thus
\be ds^2== \alpha' \left[
{1 \over \sqrt G} \left( - H dt^2 + d {\bf x}_{\parallel}^2 \right)
+ \sqrt{G} \left( {1 \over H} d {\rm U}^2 + {\rm U}^2 d {\bf \Omega}_5^2
\right) \, \right]
\label{sugra_metrics}\ee
where $\alpha'$ is the string tension, $t,{\bf x}_{\parallel}$ are 4-d
coordinate of our space-time, ${\bf \Omega}_5$ is the 5-d solid angle
of the 5-d sphere,
$G \equiv {g^2_{\rm eff} \over {\rm U}^4}$ 
%\nonumber \\
\be
H &\equiv& 1 - { {\rm U}_0^4 \over {\rm U}^4 } 
\hskip1cm {\rm U}_0^4 = {2^7 \pi^4 \over 3 } g^4_{\rm eff} \, 
{\mu \over N^2}
 \, .
%\label{harmonicftn}
\ee
where $g^2_{\rm eff}=g^2 N_c$ is the 't Hooft coupling. 
The parameter $\mu$ is interpreted as the free energy density on the near
extremal D3-brane, hence, $\mu \approx (4 \pi^2 /45) N^2 T^4$. 
So the metrics without the black hole, $T=0$ corresponds to $H=1$.

\begin{figure}[h]
 \centering 
\includegraphics[width=5.5cm]{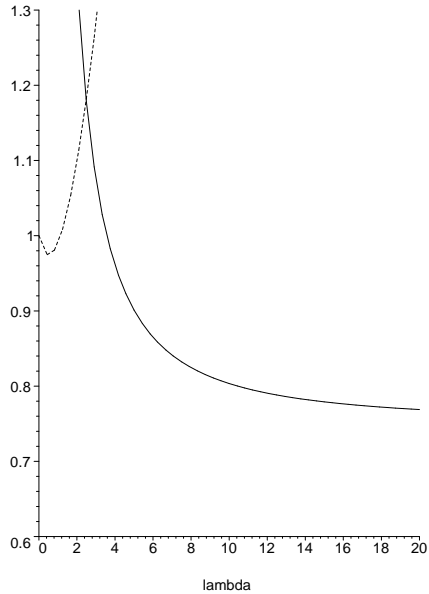}
\caption{ \label{fig_weak_strong}
 Comparison of the weak (5 terms)
 and  strong (2 terms) coupling series, for the \N=4
theory. The plot shows the pressure p (in units of its to
 Stephan-Boltzmann
value) versus the coupling $\lambda=g^2N_c$. 
At $\lambda=2-6$ both expansions seem to fail; but
presumably the true line just smoothly interpolates. In QCD at $T\sim
 1.5 T_c$ the effective coupling is at its maximum which corresponds
 to
$\lambda \sim 10-20$, which seem to be firmly on the strong coupling side.
}
\end{figure}

 The thermodynamics of this is discussed extensively 
in \cite{thermo} with the conclusion that
in the strong  coupling limit $\epsilon=3p$ reaches 3/4 of
 its
Stephan-Boltzmann value. The corrections are $O(1/\lambda^{3/2})$.
   It is very instructive to
 compare weak and strong coupling results, which we do in 
Fig.\ref{fig_weak_strong}. The pressure, shown in Fig.(a),
presumably show a monotonous decrease from 1 to 3/4, but both
weak and strong coupling results seem to be quite wrong for
$\lambda=g^2N_c=1..5$.

%%%%%%%%%%%%%%%%%%%%%%%%%%%%%%%%%%%%%%%%%%%%%%%%%%%%%%%%%%%%%%%%%%%%%%
%%%%%%%%%%%%%%%%%%%%%%%%%%%%%%%%%%%%%%%%%%%%%%%%%%%%%%%%%%%%%%%%%%%%%%%

The Maldacena's  duality 
\cite{Maldacena}, known also as Anti-deSitter AdS/CFT correspondence,
 has opened a way to study this {\em strong coupling limit} 
using classical gravity. At finite $T$ it was recently
actively discussed, for Debye screening \cite{Rey_etal},
bulk  thermodynamics \cite{thermo} and kinetics \cite{PSS}.
Although the thermodynamical quantities are only modified by an
overall factor of 3/4 in comparison to the black-body limit,
kinetics is changed dramatically.

 In our second  paper 
\cite{SZ_cft} related with bound states in QGP-like phases
we  show that in this regime the matter
is made of very deeply bound binary composites, in which the supercritical
Coulomb can be balanced by centrifugal force.
a re-summation of a class of diagrams, in vacuum and at finite $T$. 
Specific towers of such bound states can be considered as a
continuation
of Fig.\ref{fig_masses_T} to the left, toward stronger and stronger coupling.

 ${ N}=4$ super-Yang-Mills (SYM) is the most famous example of 
a Conformal Field Theory (CFT) in 4 dimensions. This theory has zero 
beta function and a non-running coupling constant, which can be continuously
changed from weak to strong. Unlike QED or QCD where for a critical
coupling $\alpha\approx 1$ there is vacuum rearrangement, the CFT is
believed to remain in the same Coulomb (plasma-like) phase for all couplings. 
Thus, it provides an interesting theoretical laboratory for understanding
properties of a strongly coupled Quark-Gluon Plasma (QGP) in QCD, which occurs
at $T\approx T_c$ as discussed in our previous paper \cite{SZ_newqgp}.

A key breakthrough in understanding the strong coupling regime was the
AdS/CFT correspondence conjectured by Maldacena~\cite{Maldacena}.
The conjecture has turned the intricacies of strong coupling
gauge theories to a classical problem in gravity albeit in 10
dimensions.
For instance, the static potential between a heavy quark
and antiquark  derived from the asymptotic of an elongated temporal
Wilson loop, follows from a minimal surface (classical string) 
between the quarks stretched by gravity (metric of the AdS
space) as depicted in Fig. 1a. The result is a modified Coulomb's 
law ($\lambda\gg 1$)~\cite{Maldacena}.

\be \label{eqn_new_Coulomb}
V(L)= -{4\pi^2  \over \Gamma(1/4)^4 }{\sqrt{\lambda} \over  L}
\label{coulomb}
\ee
The numerical coefficient in the first bracket is 0.228.
The latter will be compared to the result from a diagrammatic
re-summation below.

\begin{figure}[h!]
\centering
\includegraphics[width=8cm]{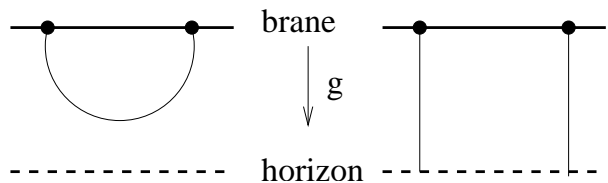}
\vskip .3cm
\caption{\label{fig_pbrane_Debye} 
Two types of solutions describing the potential between two static charges
(large dots) in the ordinary 4-d space (on the D3 brane). 
The string originating from them
can either connect them (a) or not (b). In both cases the string
is deflected by a background metric (the gravity force indicated by the
arrow marked g) downward, along the 5-th coordinate. After the string
touches the black hole horizon (b) a Debye screening of the interaction takes place.  
}
\end{figure}

The case of the non-zero temperature  is represented in the AdS
space by the occurrence of a black-hole in the 5th dimension,
whereby its Schwartzschild radius is identified with the inverse
temperature. When the string between charges
extends all the way to the black hole as shown in Fig. 1b, 
the heavy quark potential is totally screened for
a Debye radius of order $1/T$~\cite{Rey_etal}. This is to be contrasted with
$\sqrt{\lambda}T$ expected in the weak coupling limit for the electric
modes, and $\lambda\,T$ for the magnetic modes.

{ The main puzzle} related with all strong coupling results whether it 
is the free energy, the viscosity, or the resonance frequencies,
is their independence on the coupling $\lambda$ in strong coupling.
We recall that the interaction between the (quasi) particles
such as (\ref{eqn_new_Coulomb}) is proportional to $\sqrt{\lambda}$. 
In a naive picture of a quasiparticle plasma,  familiar from the
weak coupling limit, one would expect the interaction terms to show
up in the free energy. 

\subsection{Summing the ladder diagrams}

The main objective of our paper \cite{SZ_cft}
was to explain these puzzles,
but before that one should
 understand the dynamical picture behind
the modified Coulomb law, and its Debye-screened form at finite $T$ and
strong coupling. For that, one can identify  in the gauge theory a set
of diagrams whose re-summation can reproduce the parametric
features of the above mentioned strong coupling results. As a result,
we learn an important lesson: in the strong coupling regime even the
static charges communicate with each other via high frequency
gluons and scalars,  propagating with an effective super-luminal velocity
$v\approx \lambda^{1/4}\gg 1$. 

 The  
observation that  potential-type diagrams in Feynman gauge  reproduce
the strong coupling regime has been made  by Semenoff and collaborators
\cite{zarembo}, who have shown that the ladder re-summation works for the
circular Wilson loop and qualitatively explains the modified Coulomb law
at strong coupling and zero temperature.

Intuitively, the reason for a potential-like  regime stems from the
fact  that for $\lambda\gg 1$ the time between subsequent exchange of 
quanta is very short. The cost of repulsive Coulomb energy becomes 
prohibitively large at strong coupling, forcing both charges to almost 
simultaneously change their colors, keeping them oriented in mutually 
the most attractive positions. These interactions are naturally ordered 
in time, justifying the use of ladder-type approximations.

Let me start by reminding the
reader of the (Euclidean) derivation of the
standard Coulomb's law between two attractive and {abelian} 
static charges.  In the first quantized form  in Feynman gauge
one simply gets it from a 00-component of the photon (gluon) propagator

\be
V(L)\approx -{\lambda \over 4\pi^2}   \int_{-\infty}^{+\infty} \,{dt 
\over t^2+L^2}
\label{cou1}
\ee
where $t$ is the relative time separation between 
the two charges on their world lines. In the abelian case whether
at strong or weak coupling, the interaction takes place at {\bf all}
time virtualities resulting into the standard  {instantaneous} 
Coulomb interaction with $V(L)\approx -\lambda/L$. The non-abelian 
modified Coulomb's law (\ref{coulomb}) is seen to follow from 
the abelian Coulomb's law (\ref{cou1}) whereby the relative time 
interval is much shorter and of order $L/\sqrt{\lambda}\rightarrow 0$.

\begin{figure}
\centering
\includegraphics[width=12cm]{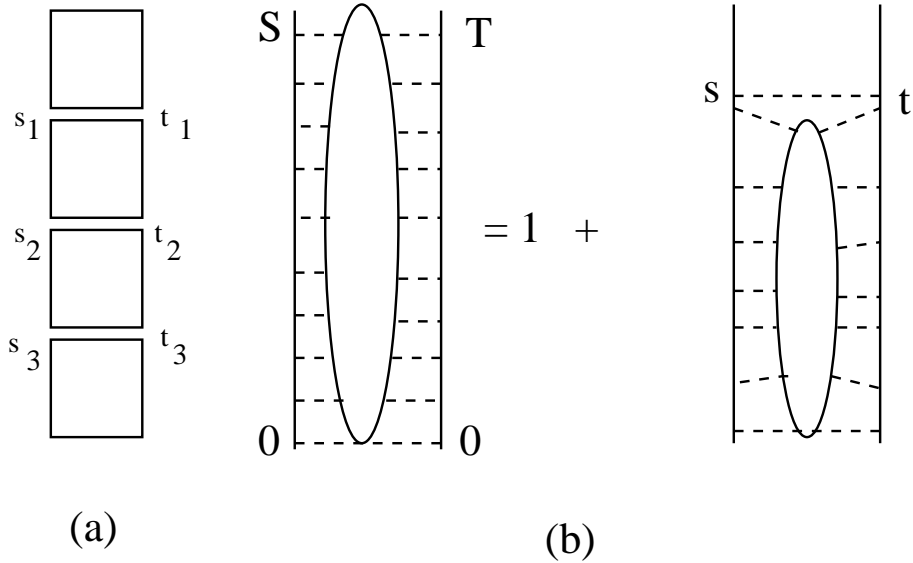}
\caption{\label{fig_ladders} 
(a) The color structure of ladder diagrams in the large-$N_c$
limit: each square is a different color trace, bringing the factor
$N_c$.
 The time goes vertically, and the planarity condition
 enforces strict time ordering, $s_1>s_2>s_3...$,
$t_1>t_2>t_3...$. (b) Schematic representation of the Bethe-Salpeter 
equation (\ref{eqn_BS}) summing ladders.
}
\end{figure}

We start by reminding the reader of some results established
in~\cite{zarembo} at zero temperature. At large $N_c$ 
(number of colors), the  diagrams can be viewed
as t'Hooft diagrams. An example of a ladder diagram
is shown in Fig.~\ref{fig_ladders}a, where the rungs can be either
gluons and scalars, as both are in the adjoint representation.
The first lesson is that each rung contributes a factor $N_c$,
which however only comes from planar diagrams. It means, that
in contrast to the Abelian theory, the time ordering should be strictly
enforced,  $s_1>s_2>s_3...$ and $t_1>t_2>t_3...$, which will be 
important below.

In Fig.~\ref{fig_ladders}b we schematically
depict  the Bethe-Salpeter equation. The oval connected with
the two Wilson lines by multiple gluon/scalar lines
is  the re-summed Bethe-Salpeter kernel $\Gamma(s,t)$, describing
the evolution from time zero to times $s,t$ at two lines. It 
satisfies the following integral equation

\be \label{eqn_BS}
\Gamma(S,T)=1+{\lambda \over 4\pi^2}\int_0^S ds \int_0^T dt{1\over
    (s-t)^2+L^2} \Gamma(s,t) \ee
which provides  the re-summation of all the
ladder diagram. $L$ is the distance between two charges, and 
the first factor under the integral is the (Euclidean) propagator for
one extra gluon/scalar added to the ladder. The kernel obviously
satisfies the boundary condition $\Gamma (S,0) =\Gamma(0,T)=1$.
If the equation is solved, the ladder-generated potential follows from
$
V_{\rm lad}(L) 
=-\lim_{T\to{+\infty}}{\frac 1T \Gamma\, (T,T)}\,\,,
\label{0a}
$.

In weak coupling $\Gamma\approx 1$  and the integral on the rhs is
easily taken, resulting in 
\be
\Gamma (S,T) \approx 1+\frac{\lambda}{8\pi}\,\frac{S+T}L
\label{00a}
\ee
which results into the standard Coulomb's law.
Note that in this case the typical relative time difference
between emission and absorption of a quantum $|t-s|\approx L$, so one can
say that virtual quanta travel at a speed $v\approx 1$.

For solving it at any coupling, it is convenient
to switch to the differential equation

\be
\frac{\partial^2\Gamma}{\partial S\,\partial T} =
\,\frac{\lambda/4\pi^2}{(S-T)^2+L^2}
\Gamma (S,T)\,\,\,.
\label{1a}
\ee
and change variables to
$x=(S-T)/L$ and $y=(S+T)/L$ through
\be
\Gamma (x,y) =\sum_{m}\,{\bf C}_m \gamma_m (x)\,e^{\omega_m y/2}
\label{2a}
\ee
with the corresponding boundary condition $\Gamma (x,|x|)=1$. The
dependence of the kernel $\Gamma$ on the relative times $x$ follows
from the differential equation

\be
\left(-\frac{d^2}{dx^2} -
\frac{\lambda/4\pi^2}{x^2+1} \right)
\,\gamma_m (x) = -\frac {\omega_m^2}{4}\,\gamma^m (x)
\label{3a}
\ee
For large $\lambda$ the dominant part of the potential in (\ref{3a})
is from {\bf small} relative times $x$ resulting into a harmonic
equation~\cite{zarembo}

\be
&&\left(-\frac{d^2}{dx^2} +\frac 12
({\lambda/4\pi^2})\,x^2 \right)
\,\gamma_m (x) \nonumber\\
&&= -\frac 14 \left({\omega_m}^2-{\lambda}/{\pi^2}\right)\,
\gamma_m (x)\,\,.
\label{4a}
\ee
This shows that the sum of the ladders grow exponentially. At large 
times $T$,  the kernel is dominated by the lowest harmonic mode of
(\ref{4a}). For large times $S\approx T$ that is small $x$ and large 
$y$ 

\be
\Gamma (x,y)\approx {\bf C}_0\,e^{-\sqrt{\lambda}\,x^2/4\pi}\,
e^{\sqrt{\lambda}\,y/2\pi}\,\,.
\label{5a}
\ee
From (\ref{0a}) it follows that
in the strong coupling limit the ladder generated potential
is 

\be V_{\rm lad}(L)= -\frac{\sqrt{\lambda}/\pi}L \ee which 
has the same parametric form  as the one derived from the
AdS/CFT correspondence except for the
overall coefficient\footnote{ Note that the difference
is not so large,  since $1/\pi=0.318$ is larger than the exact value  
0.228 but about 30\%. So additional screening by about 1/3 is needed 
to get it right. This is of course in the left out higher order
diagrams.}.

The results of ref.  \cite{zarembo} discussed above indicate
that summing ladders get some vacuum physics but {\bf not all} 
since the overall coefficient is not reproduced exactly. The same conclusion follows from
the fact that the expectation values of Wilson lines are gauge invariant, while
the ladder diagrams are not. Therefore, some non-ladder
diagrams must be equally important and should be included.

Let me skip discussion of any higher order diagrams,  summarizing
what we have learned about the ladder re-summation which are worth
stressing. In these diagrams the
 transverse jumps of the interaction points
 would be as small as the time steps
namely,
\be \delta {\bf x}_t \sim
L/\lambda^{1/4}
\ee

\begin{wrapfigure}{l}{5.cm}
\begin{minipage}{4.cm}
\centering
\includegraphics[width=4cm]{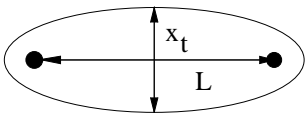}
\caption{\label{fig_planedistribution}
Distribution of the interaction vertices in space. The black
circles are static charges. 
}
\end{minipage}
\end{wrapfigure}
Thus the $d{\bf x}$ integration yields a suppression factor of
$(1/\lambda^{1/4})^4\approx 1/\lambda$.  Since the $ds\,dt$ 
integrations bring about a suppression $(1/\lambda^{1/4})^2\approx
1/\sqrt{\lambda}$, it follows that it is of order
$\sqrt{\lambda}$. This is precisely {\bf the same order}
as the modified potential discussed in the previous subsection.
So, such higher order diagrams are neither
larger nor smaller than the ladder ones we discussed earlier.
What we retain from this discussion, 
is that all extra quanta prefer to be in
an ellipsoidal region of space as shown
in Fig.~\ref{fig_planedistribution}.
The transverse sizes ${\bf x}_t\approx 
\delta t\approx L/\lambda^{1/4}$.
In QCD and other confining theories we known that gluons
make a string between two charges with a constant width and tension:
attempts to derive it from re-summed diagrams continue. 
In CFT under consideration
the width in transverse direction must be proportional to $L$
since the conformal symmetry prohibits any other dimensional scale
to be developed. Still, in strong coupling
the ellipsoid is very elongated ${\bf x}_t\approx L/\lambda^{1/4}\ll L$:
this is what we meant by a ``quasi-string'' regime in the title of
this subsection.

\subsection{Bound states of the quasiparticles}

The simplest bound state problem in our case is that of a scalar
in the presence of an infinitely heavy source with compensating color
charge. The latter acts as an overall attractive Coulomb potential
$V$ (the effects of screening will be discussed below). In strong
coupling $V$ acts on the accompanying relativistic gluino
quasi-instantaneously. For a spherically symmetric potential,
the Klein-Gordon equation  reads
$$
-{d^2   \over dr^2}\chi_l=\left[(E-V)^2-m^2-{\tl^2\over
r^2}\right]\chi_l
$$
with the  wave function $\phi=Y_{lm}\chi_l(r)/r$ and the orbital
quantum number $\tl^2=l(l+1)$\footnote{ For a semiclassical analysis
at small $l$ one can
use the Langer prescription $l(l+1)\rightarrow (l+1/2)^2$, which is
known to yield semiclassically stable S-states. However we will only
need large $l$.}. For a
Coulomb-like potential, the equation is exactly solvable in terms of
hyper-geometric functions,
from which the quantized energy levels are
 \be \label{eqn_spectrum_sctrongcoupling}
E_{nl}=\pm m \left[1+
\left({C\over n+1/2+\sqrt{\tl^2-C^2} } \right)^2\right]^{-1/2}\,\,.
\ee

\begin{wrapfigure}{l}{6.cm}
\begin{minipage}{5.cm}
\centering
\includegraphics[width=5cm]{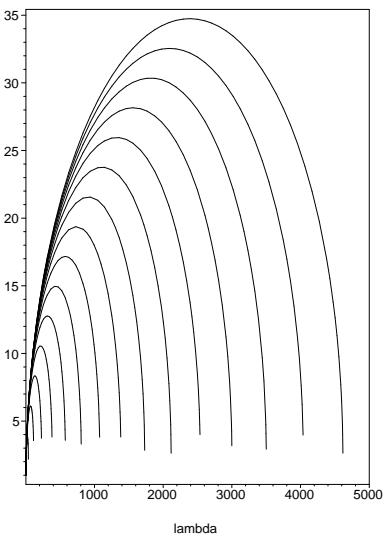}
%\vskip .4cm
%\includegraphics[width=5cm]{wkb02.eps}
\caption{\label{fig_wkb} 
The  spectrum of the states versus the 'tHooft coupling constant $\lambda$.
for the levels with fixed $n_r=0$ and the orbital momentum $l=1..15$.
One can see that there are light bound states at any coupling. 
%, (b) levels with fixed
%$l=10$ and $n=0..5$. 
}
\end{minipage}
\end{wrapfigure}
In weak coupling $C=g^2N=\lambda$ is small and the bound states energies
are close to $\pm m$. Specifically~\footnote{The fact that only the
combination $n+l$ appears, i.e. principle  quantum number, is a
consequence of the known Coulomb degeneracy. This is no longer
the case in the relativistic case.}
\be 
E_{nl}= m -{C^2 m \over 2(n+l+1)^2}\,\,, 
\ee
which is the known Balmer formula.
All of that, including the expression above, was known since 
1930's.

New view on this formula,
in the (opposite) strong coupling limit,
 gives the following. If the Coulomb law coefficient is large
$C=(4\pi^2 / \Gamma(1/4)^4 )\sqrt{\lambda} \gg 1$,
the quantized energies are imaginary {\em unless the square
root gets balanced by a sufficiently large angular momentum}. 
 In this regime, one may ignore the $1$ in 
(\ref{eqn_spectrum_sctrongcoupling}) and obtain the {\bf equi -distant}
spectrum of deeply bound states

\be
E_{nl}\approx {m\over C}\left[ 
(n+1/2)+\left((l+1/2)^2-C^2\right)^{1/2} \right] 
\label{wkb}
\ee

Since the mass is related to the thermal loop with $m\approx
\sqrt{\lambda}$, the ratio $m/C\approx T\lambda^0$ is proportional to 
temperature but {\bf independent} of the coupling constant.
More details on the energy dependence on the coupling
constant are shown in Fig.~\ref{fig_wkb}
for different values of the orbital quantum number $l$ (a) and 
radial quantum number $n$ (b). All lines end when the Coulomb
attraction is able to overcome the centrifugal repulsion.

After we established the spectrum, let us see its consequences for
{\em thermodynamics and kinetics at strong coupling}.
 Normally all kinetic
quantities are related with scattering cross sections, and the absence
of {\bf any} coupling is a priori implausible physically. In general,
thermodynamical quantities count degrees of freedoms and one may think 
that the coupling constant may indeed be absent. However, the effective 
masses of the quasiparticles and all pairwise interactions are still 
proportional to $\sqrt{\lambda}$  and in general $\lambda^\alpha$ with
$\alpha>0$. If so, the quasiparticle contributions to
the statistical sum should be exponentially small at strong coupling,
of the order of $ e^{-\lambda^\alpha}\ll 1$, and one will have to conclude that
some other degrees of freedom are at play.

Indeed there are new degrees of freedom at play in the strong coupling
regime of the thermal Coulomb phase. As we have
shown above, for large $\lambda$ there are light (deeply bound) binary
composites, with masses of order $T$ irrespective of how strong is
$\lambda$. The composites are almost point like with 
 thermal sizes of the order of $1/T$, which 
readily explains the liquid-like kinetic behavior. The thermal Coulomb
phase is a liquid of such composites. The composites are light and should dominate the 
long-distance behavior of all the finite temperature Euclidean correlators.

We start by explaining the leading contribution to free energy from
the gauge theory point as a liquid of composites. The factor of $N_c^2$
in front of the free energy and the viscosity for instance, follows from
the fact that {\bf all} our composites are in the adjoint
representation. Indeed, for fundamental charges the composites are
meson-like (color neutral) with all color factors absorbed in the
coupling constant. For adjoint charges, there is in addition {\bf two}
spectator charges~\footnote{For example, a composite can be initially made of
blue-red and red-green colors, with red rapidly changing inside the
ladder.} that do not participate into the binding. The Coulomb phase
is not confining. Thus the number of light and Coulomb bound composites
is $N_c^2$.

The astute reader may raise the question that by now our arguments
are totally circular: we have started with $N_c^2$ massless relativistic
states and we have returned to $N_c^2$ light composites. So what is the
big deal? Well, the big deal is that we are in strong coupling, and in
fact the degrees of freedom completely reorganized themselves in
composites, for otherwise they become infinitely heavy thermally
and decouple. This is how Coulomb's law negotiates its deeds in a system
with very large charges.

The composites carry large angular momentum in strong coupling, 
i.e. $l\approx \sqrt{\lambda}$. Their weight contribution to the
partition function is of the order of

\be 
\int_{l_{min}}^{l_{max}} dl^2=l_{max}^2-l_{min}^2\approx \lambda^0
\label{lw}
\ee
which is independent of $\lambda$, since the WKB orbits are stable
only for $l\approx (\sqrt{\lambda} +1/\sqrt{\lambda})$. The ensuing
thermodynamical sum over the radial quantum number $n$ is
independent of $\lambda$, leading to a free energy ${\bf F} ={\bf C}
N_c^2\,T$ in agreement with strong coupling result. For weak coupling ${\bf C}=1$
while for strong coupling ${\bf C}=3/4$.

Let us now summarize this section.
We have exploited the fact that ${ N}=4$ SYM is in a Coulomb
phase at all couplings, to argue that in strong coupling (Maldacena
regime) color charges only communicate over very short period of times
$t\approx L/\lambda^{1/4}$ for a fixed separation $L$. This physical
observation is enough to show why ladder-like diagrams in the gauge
theory reproduce the modified Coulomb's law obtained by the AdS/CFT 
correspondence using classical gravity. 

These observations are generic and suggest that in the gauge theory the
modified Coulomb potential applies equally well to relativistic and
non-relativistic charges. Indeed, since the relativistic particles move
with velocity $v\approx 1$, the color reordering encoded in the 
modified Coulomb potential goes even faster through a virtual quantum 
exchange with velocity $v\approx \lambda^{1/4}\gg 1$. At strong
coupling, the color charge is so large that color rearrangement is
so prohibitive unless it is carried instantaneously. This is the only 
way Coulomb's law could budget its energy. The same observations extend 
to finite temperature
where we have shown that the modified Coulomb potential acquires a
screening length of the order of $1/T$ irrespective of how strong is
the coupling. 

We have analyzed the effects of a (supercritical) Coulomb field on the 
motion of colored relativistic particles. Bound states form whenever 
the squared Coulomb potential balances the effects of centrifugation.
 Rather unexpectedly, we have
found that even though the trajectory of any particular Coulomb bound
state depends critically on the coupling $\lambda$, their average
density remains about $\lambda$-independent constant. We hope,
this explains
puzzling results obtained using the string theory.

\section{Strongly coupled  trapped atoms}

\begin{wrapfigure}{l}{5.cm}
\begin{minipage}{4.cm}
\centering
\vskip -.4cm
\includegraphics[width=4cm]{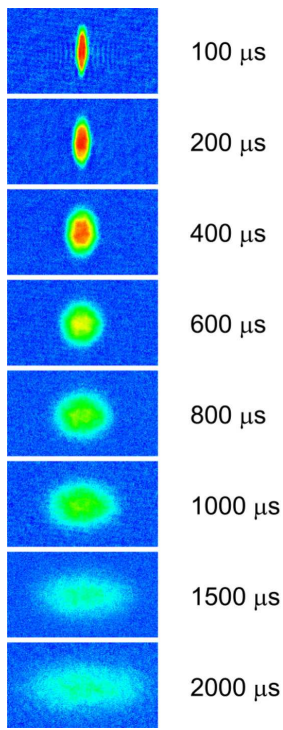}
\caption{\label{fig_atoms} 
Hydrodynamical elliptic flow of trapped $Li^6$ at strong coupling,
 from \cite{Li6}. 
}
\end{minipage}
\end{wrapfigure}

Experimental studies of strongly coupled many-body problems, in
which
the binary scattering length diverges, are recently done
with
trapped atomic $Li$ atoms.  The scattering length can be tuned
 till practically infinite values, plus or minus, by applying
a magnetic field which shifts the Feshbach resonances.

Remarkably, for its fermionic version it was indeed
found very recently
that a strong coupling leads to a hydrodynamical behavior
\cite{Li6}. The way it was demonstrated is precisely
the same ``elliptic flow'' as discussed above. One can
start with a deformed trap.  Normally the gas is so dilute
$a n^{1/3} \ll 1$ that atoms just fly away isotropically,
but when tuning to strong coupling regime is done
the expansion is anisotropic and can be described hydrodynamically. 

A number
of other spectacular
 experimental discoveries with trapped $Li^6$
 were also made later. 
 It was found \cite{atoms_capture} that
an adiabatic crossing through the resonance 
 converts nearly all atoms into very loosely bound
(but remarkably stable)  ``Cooper pairs'', which can also
Bose-condense  \cite{atoms_molec_cond} if the temperature is low enough.   

Since in heavy ion collisions the system also crosses
the zero binding lines adiabatically, various bound pairs of quarks and gluons
should also be generated this way. 
One might think that such states are pre-prepared at crossing of the
zero
binding lines, not at the critical line.

\section{Brief summary}
One of the main lessons we have learned from RHIC data is that
the QGP we found is very far from a weakly interacting gas of quasiparticles.
Although the EoS fitted by hydrodynamics agrees well with lattice data,
its transport properties is very different. So, whatever it takes,
we should learn more about more strongly coupled systems.

One consequence of stronger coupling is appearance of bound states
of quasiparticles. As  emphasized by Zahed and myself \cite{SZ_newqgp},
near the zero binding lines one may indeed expect large rescattering.
The experience with trapped cold atoms support this mechanism of hydro
very strongly.

Another super-strongly interacting system is \N=4 SUSY YM theory:
it also leads to hydrodynamical behavior and small viscosity.
And again, the bound states seem to be present 
\cite{SZ_cft} and may play a key role.

\end{document}